%% file: jackson_manuscript_2013Nov20.tex
\documentclass[aps,apl,10]{emulateapj-rtx4}

\slugcomment{{\sc Received by ApJ 2013 August 6; accepted by ApJ 2013 October 14; published by ApJ 2013} }

\usepackage{lscape}
\usepackage{graphicx}
\usepackage{amsmath}
\usepackage{natbib}
\newcommand{\overallnumbins}{200\ }
\newcommand{\exampleKIC}{KIC 7582691}

\newcommand{\numnophotoshift}{four}
\newcommand{\tess}{\emph{TESS}}
\newcommand{\kepler}{\emph{Kepler}}
\begin{document}

\title{A Survey for Very Short-Period Planets in the Kepler Data}
\author{Brian Jackson\altaffilmark{*}}

\author{Christopher C. Stark}
\affil{Carnegie Institution for Science, 5241 Broad Branch Road, NW, Washington, DC 20015, USA}
\author{Elisabeth R. Adams}
\affil{Planetary Science Institute, 1700 E. Ft. Lowell, Suite 106, Tucson, AZ 85719, USA}

\author{John Chambers}
\affil{Carnegie Institution for Science, 5241 Broad Branch Road, NW, Washington, DC 20015, USA}
\author{Drake Deming}
\affil{Department of Astronomy, University of Maryland at College Park, College Park, MD 20742, USA}
\altaffiltext{*}{bjackson@dtm.ciw.edu}

\begin{abstract}
We conducted a search for very short-period transiting objects in the publicly available \kepler~dataset. Our preliminary survey has revealed \numnophotoshift planetary candidates, all with orbital periods less than twelve hours. We have analyzed the data for these candidates using photometric models that include transit light curves, ellipsoidal variations, and secondary eclipses to constrain the candidates' radii, masses, and effective temperatures. Even with masses of only a few Earth masses, the candidates' short periods mean they may induce stellar radial velocity signals (a few m/s) detectable by currently operating facilities. The origins of such short-period planets are unclear, but we discuss the possibility that they may be the remnants of disrupted hot Jupiters. Whatever their origins, if confirmed as planets, these candidates would be among the shortest-period planets ever discovered. Such planets would be particularly amenable to discovery by the planned \tess~mission.
\end{abstract}


\maketitle

\section{Introduction}
\indent Almost every new discovery in extrasolar planetary (exoplanetary) astronomy challenges a previously-held truth about the origins and natures of planets. In particular, the discoveries of hundreds of exoplanets orbiting within 0.1 AU of their host stars with orbital periods of a few days (close-in planets) have defied conventional planet formation and evolution models based on our solar system. Tidal interactions between these planets and their host stars can influence the orbits and the stellar rotation state. In addition, tidal interactions have probably played a role in the planets’ origins (e.g. \citealp{2007ApJ...669.1298F}).

\indent The tide raised on the star can induce orbital decay long after an initial non-zero orbital eccentricity has damped to small values, as long as the stellar rotation remains unsynchronized with the planet's orbital motion. In such a case, the planet's orbital stability depends entirely on the total (star's rotational plus the orbital) angular momentum \citep{1973ApJ...180..307C}, and for planetary systems with orbits well-aligned with the stellar equator, \citet{2009ApJ...692L...9L} gave the following expression for the critical angular momentum required for stability $L_\mathrm{stab}$:
\begin{equation}
L_\mathrm{stab} = 4 \bigg[\frac{G^2}{27}\frac{M_\star^3 M_p^3}{M_\star + M_\mathrm{p}} \left(C_\star + C_\mathrm{p}\right)\bigg]^{1/4}
\label{eqn:thresh_ang_mom}
\end{equation}
where $G$ is Newton's gravitational constant, $M_\mathrm{\star/p}$ the stellar/planetary mass, and $C_\mathrm{\star/p}$ the stellar/planetary moment of inertia. 

\begin{figure}
\includegraphics[width=0.48\textwidth]{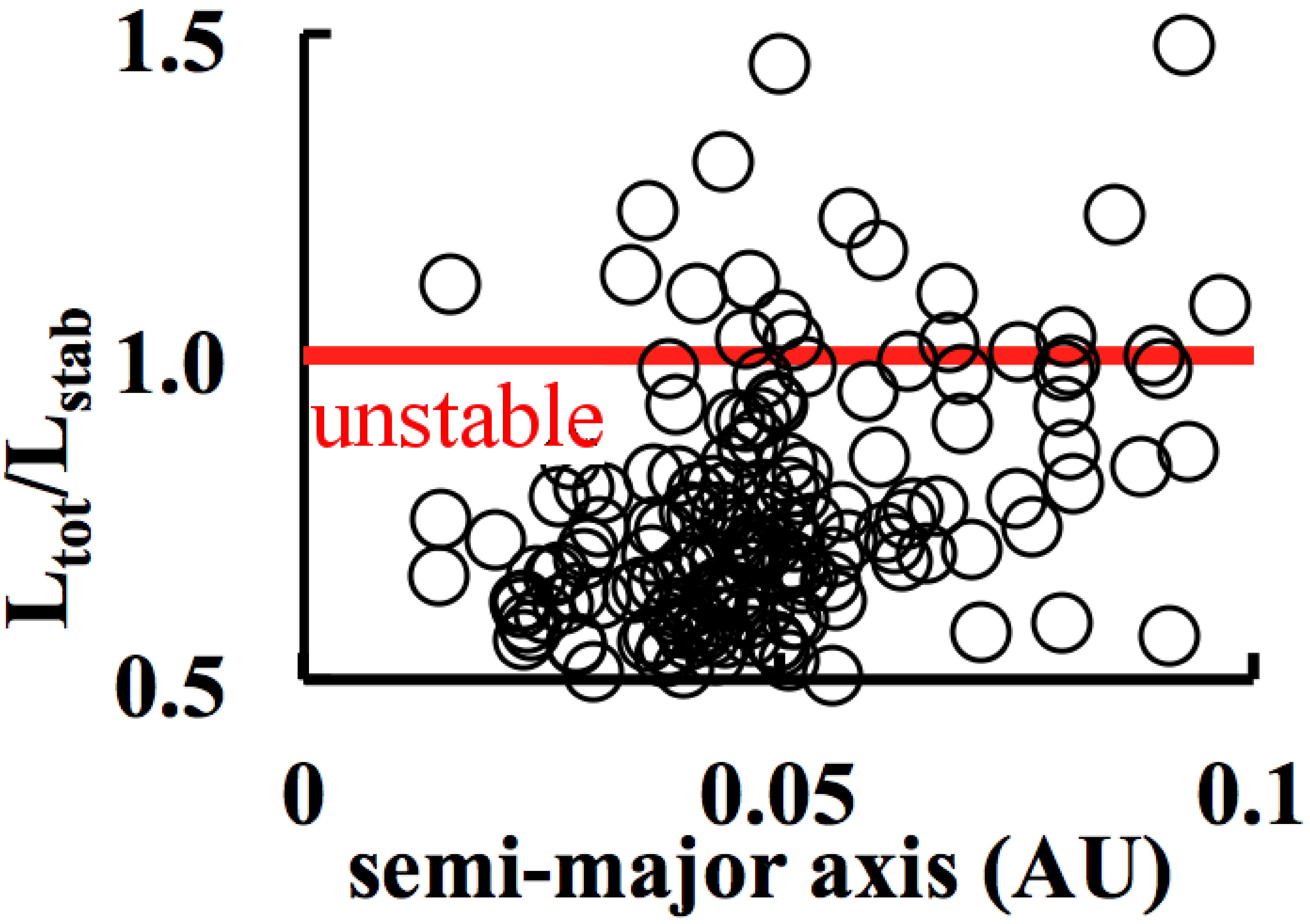}
\caption{Total angular momentum $L_\mathrm{tot}$ (orbital + rotational) for many hot-Jupiter planetary systems, normalized to the threshold angular momentum required for the planet to be stable against tidal decay $L_\mathrm{stab}$. Data taken from exoplanets.org on 2012 April 23. Where unavailable, stellar rotation periods are assumed to be 30 days.}
\label{fig:ang_mom}
\end{figure}

\indent Figure \ref{fig:ang_mom} shows estimated total angular momenta for many hot-Jupiter planetary systems $L_\mathrm{tot}$, normalized to the threshold value $L_\mathrm{stab}$ for each system. For the figure, we used data from exoplanet.eu. Where stellar rotation rates weren't available, we assumed a Sun-like rotation rate, and we assumed zero stellar obliquity. Unstable systems have $L_\mathrm{tot}/L_\mathrm{stab} < 1$ and lie below the ``unstable'' line. In fact, even some of the systems above the line may be unstable since the stars continually shed angular momentum through stellar winds \citep{1972ApJ...171..565S}. As the figure shows, most hot Jupiters are formally unstable against tidal decay.

\indent It is possible that orbital decay proceeds slowly enough that the formally unstable planets are actually safe for longer than the main-sequence lifetime of their host stars, as may be the case for the relatively far-out planet 51 Peg b \citep{1996ApJ...470.1187R}. (Close-in planets undoubtedly are accreted once their host stars leave the main sequence and expand -- e.g. \citealp{2009ApJ...700..832C}.) The physics of tidal dissipation, which determines the rate of evolution, remains poorly understood, and estimated tidal decay timescales span orders of magnitude \citep{2007ApJ...661.1180O}. However, planets much closer in than 51 Peg b have been found, and \citet{2009ApJ...698.1357J} showed the distribution of orbital semi-major axes for close-in planets is consistent with complete orbital decay and accretion of the planets during the main sequence. The bottom line is that the long-term orbital stability of short-period exoplanets is in question, and it is entirely plausible that close-in planets spiral into their host stars before the stars leave the main sequence. 

\indent Motivated by these considerations, we conducted a survey for such short-period planets using data from the \kepler~mission. \kepler~has already revolutionized exoplanetary astronomy, and the exquisite precision and long baseline of its photometric data make it well-suited for finding very short-period transiting planets. However, given the mission's prime goal of finding habitable (and thus, longer-period) planets, its transit search algorithm focuses on periods $>$ 0.5 days \citep{2013ApJS..204...24B} and so is not optimized to find very short-period planets (J. Jenkins, private communication, 2012). Thus, a search for planets with periods $<$ 0.5 days would uniquely exploit the large pool of publicly available \kepler~data.

\indent Some groups have begun discovering short-period objects in the \kepler~dataset. \citet{2012ApJ...752....1R} reported an object with an orbital period $\approx$ 16-hour and a variable transit light curve that may be a disintegrating rocky body. \citet{2011Natur.480..496C} claimed to have found two non-transiting Earth-sized planets with periods of 3 and 5 hours around a post main sequence (MS) B star that probably originated through interaction with the ejected stellar envelope. \citet{2013MNRAS.429.2001H} found a handful of transiting objects with periods $<$ 0.5 days; \citet{2013A&A...555A..58O} found one. Very recently, \citet{2013ApJ...774...54S} announced discovery of a planet with an 8.5 hour orbital period, Kepler-78 b (KIC 8435766).

\indent In this study, we report the results of our own preliminary search for transiting planets with periods $<$ 0.5 days. We find putative transit signals for \numnophotoshift \kepler~targets, all of which clearly pass the standard tests for \kepler~false positives and all with periods less than twelve hours. The radii for most of our candidates are also consistent with expectations for rocky planets. Ground-based radial velocity (RV) and high-spatial resolution adaptive optics (AO) observations are required to confirm these candidates as planets, but if confirmed, they will be among the shortest-period planets discovered so far. 

\indent In Section \ref{sec:data_analysis}, we detail our transit search and data analysis. In Section \ref{sec:results}, we present the results of our survey and discuss individual candidates. In Section \ref{sec:discussion_and_conclusions}, we speculate on the origins and natures of the candidates and discuss future prospects. 

\section{Data Analysis}
\label{sec:data_analysis}

\indent In this section, we describe how we conditioned and sifted the raw \kepler~data for planetary candidates. We verified that our process worked by comparing our detections to other surveys of the \kepler~data. For example, we recovered all of the eclipsing binaries with periods $\le 0.5$ days described by \citet{2011AJ....142..160S}. We removed these binaries from our list of candidates since they are not planets. We also recovered the handful of KOIs (Kepler Objects of Interest) with periods $\le 0.5$ days reported in \citet{2013ApJS..204...24B}. Our process found many of the longer period KOIs but with periods aliased to $< 0.5$ days. We also recovered the short-period candidates reported in \citet{2013MNRAS.429.2001H}, \citet{2013A&A...555A..58O}, and \citet{2013ApJ...774...54S}. 

\subsection{Data Conditioning}
\label{sec:data_cond}

\indent We retrieved the long-cadence (30-min) publicly available \kepler~data for all available targets from the MAST archive\footnote{\href{http://archive.stsci.edu/kepler/data\_search/search.php}{http://archive.stsci.edu/kepler/data\_search/search.php}}. For each target, we retrieved as many quarters as were available from among Quarters 0-11 (Q0-11). We used the raw data (called ``SAP\_FLUX'' in the FITS files), which exhibit a variety of astrophysical and instrumental variations that act as sources of noise for our analysis. 

\indent There are several techniques for mitigating these variations, and the \kepler~team has provided models for the instrumental variations\footnote{\href{http://archive.stsci.edu/kepler/cbv.html}{http://archive.stsci.edu/kepler/cbv.html}}. However, for our search, we took advantage of the fact that the transits in which we're interested have very short durations (typically tens of minutes) compared to the usual durations of interest for transit searches.

\begin{figure*}[b]
\includegraphics[width=\textwidth]{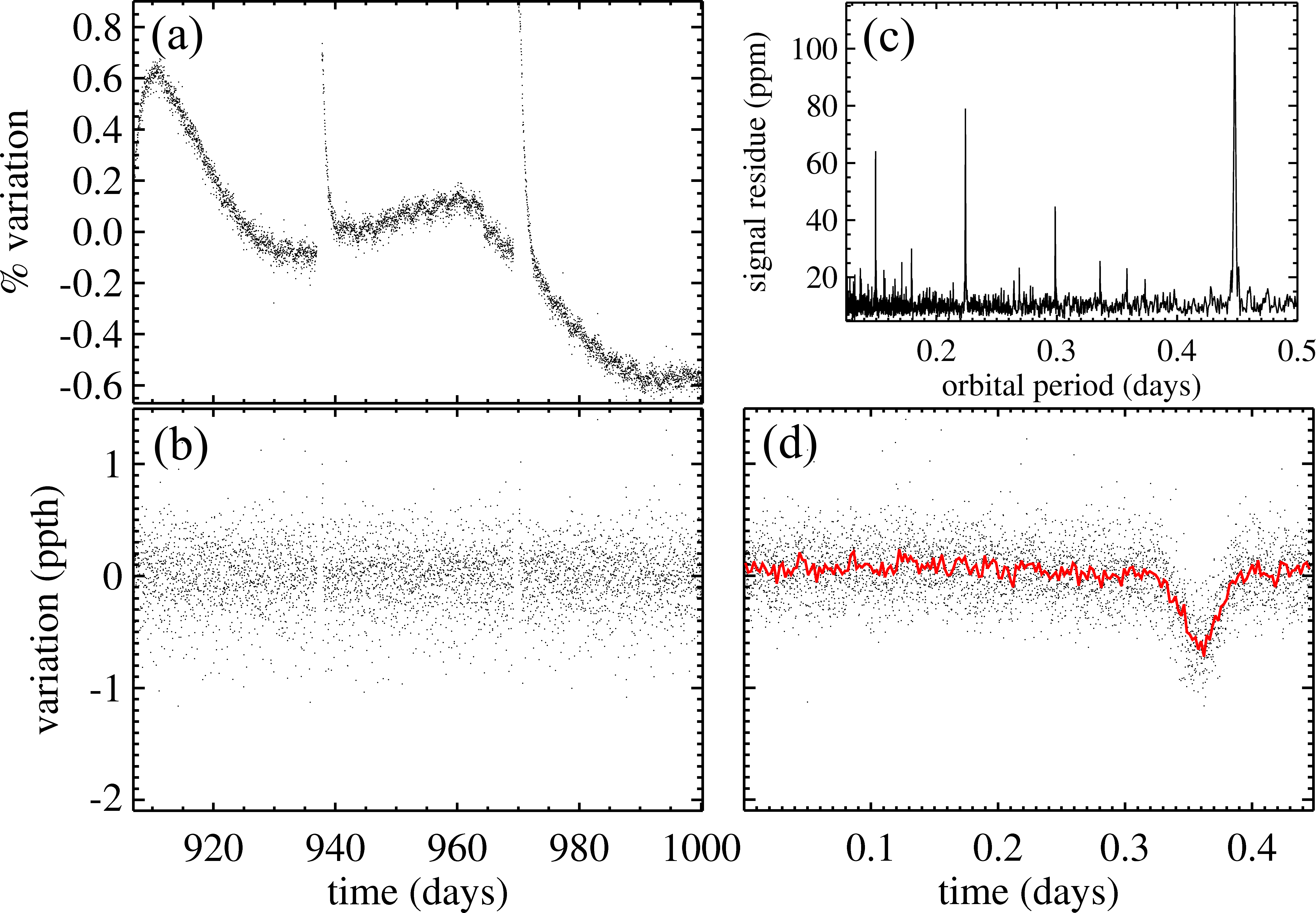}
\caption{(a) Raw \kepler~data for the KIC 10453521 candidate planetary system. (b) Detrended data in parts-per-thousand ppth. (c) EEBLS spectrum, with a peak at about 0.44 days ($\approx$ 11 hours). (d) Data from (b) folded on that 11-hour period, with variations measured in parts-per-thousand (ppth). The red line shows the folded data binned to 3-min wide bins.}
\label{fig:find_transit}
\end{figure*}

\indent To condition each quarter's observations of each target, we removed data points that are flagged as bad by the \kepler~team (with NaN or infinity in the FITS file). We subtracted the quarter's mean value from all data points and then normalized by that mean. To these mean-subtracted, mean-normalized data, we applied a mean boxcar filter with a width of 0.5 days (i.e. we calculated the mean in a window 0.5 days wide around each data point and subtracted this value from the data point). With this width, the boxcar usually contained about 20 points. Through numerical experimentation, we verified that this detrending negligibly distorts even transits with orbital periods of 0.5 days and durations longer than we searched for. Finally, we stitched together all quarters for a particular target and masked out 10-$\sigma$ outliers. \footnote{We define $\sigma$ to be the standard deviation estimated as 1.4826 $\times$ the median absolute deviation \citep{2003drea.book.....B}.} These are the data in which we searched for transits. Figure \ref{fig:find_transit} illustrates this process for one target's data. Given the typical ingress/egress times for our candidates ($\sim$ 1-min long) and \kepler's 30-min observing cadence, the transits are shallower and more v-shaped than usual planetary transits.

\indent For the candidates we found (described below), we investigated the noise properties of the detrended data using a binning test similar to those described in \citet{2011ApJ...740...33D}. We folded a candidate's detrended data on the best-fit period, binned the data into 200 bins evenly spaced in orbital phase, and subtracted each bin's mean value from all the points in that bin. (This choice of bin number was motivated by the balance between having sufficient time resolution for very short transits and producing sufficiently high signal-to-noise ratios to see the transits in the first place.) This process should remove all transits and other signals, leaving (presumably) only Gaussian residuals.

\begin{figure}
\includegraphics[width=0.48\textwidth]{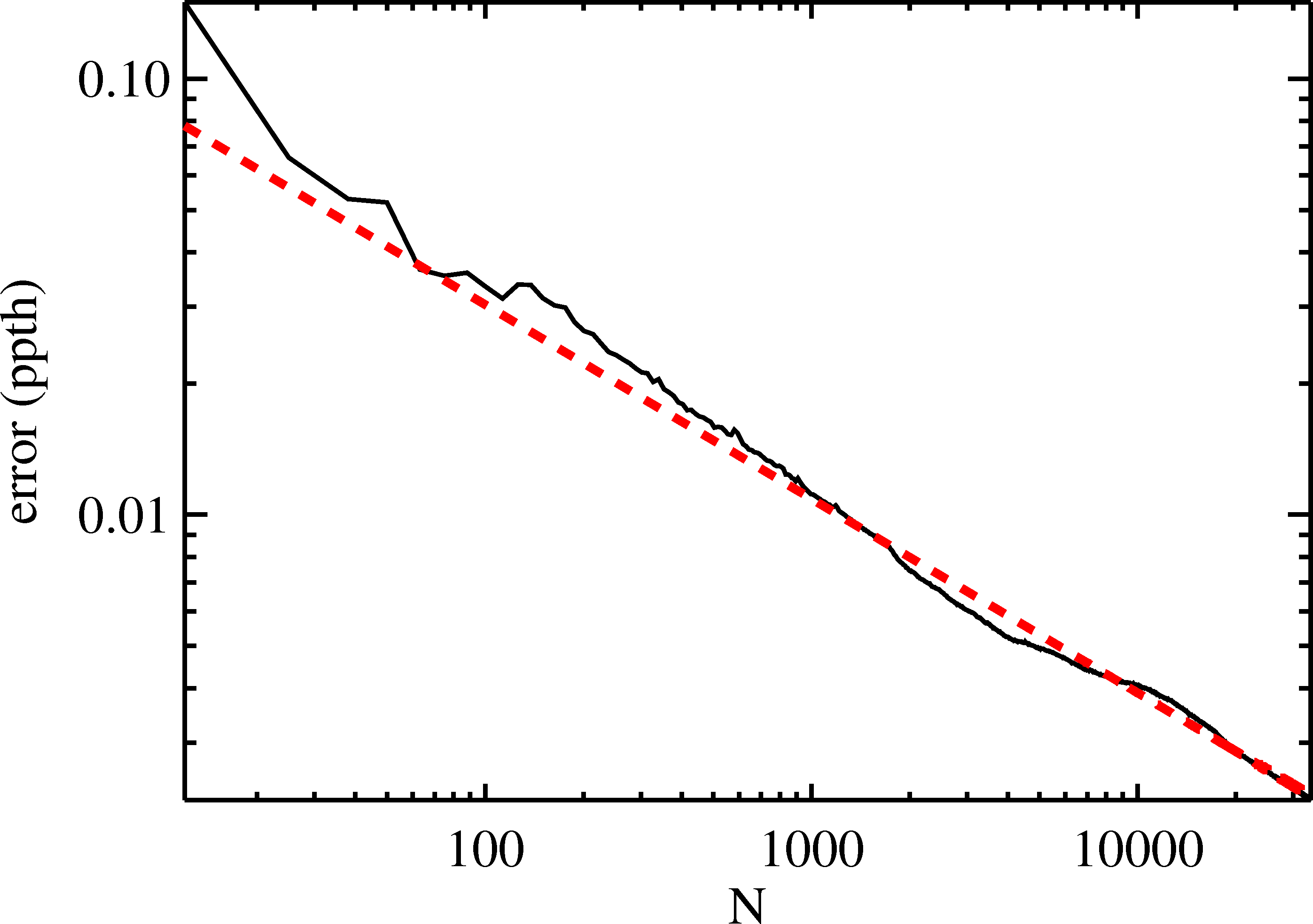}
\caption{The relationship of the standard error of the mean (y-axis, in parts-per-thousand ppth) of the data to the number of orbits folded/binned together for KIC 7582691, and the line shows the best-fit $N^{-0.48}$ relationship.}
\label{fig:KIC7582691_binning_test}
\end{figure}

\indent We then took spans of data as long as each object's orbital period and folded together a number of orbits $N_\mathrm{orb}$. Starting with $N_\mathrm{orb} = 1$, we calculated the standard error of the mean for each span of data and then averaged together the errors from all the data spans. Then, we took $N_\mathrm{orb} = 2$, then 3, and so on, until all data for an object were folded on the orbital period. The number of points folded together for a given $N_\mathrm{orb}$ is $N$, and correlated noise (if it exists) should cause the error to diminish more slowly than as $N^{-1/2}$. Figure \ref{fig:KIC7582691_binning_test} shows the results from this binning procedure for one of our targets, \exampleKIC. The slopes for the majority of our targets in $\log$(error) vs. $\log(N)$ space agree with -1/2 to within a few percent, suggesting minimal correlated noise. In any case, we used a Markov-Chain Monte-Carlo (MCMC) scheme \citep{2005AJ....129.1706F} in our model analyses, so our results do not rely on the assumption of Gaussian noise.

\subsection{Searching for Candidates}
\label{sec:searching_for_candidates}

\indent Using these data, we searched for short-period planetary transits with the EEBLS algorithm \citep{2002A&A...391..369K}\footnote{The original FORTRAN version of EEBLS is here: \href{http://www.konkoly.hu/staff/kovacs/index.html}{http://www.konkoly.hu/staff/kovacs/index.html}. We ported the algorithm to IDL, and it, along with many of the other routines we used, is available here: \href{http://www.lpl.arizona.edu/~bjackson/idl\_code/index.html}{http://www.lpl.arizona.edu/$\sim$bjackson/idl\_code/index.html}.}. Briefly, the EEBLS algorithm folds data on several trial orbital frequencies and, considering all relevant orbital phases for a given frequency, returns the best-fit transit depth at each requested frequency.

\indent We considered the orbital frequency range $(f_{min}, f_{max}) = (2, 12)$ days$^{-1}$, corresponding to periods between 2 hrs (an orbit near the surface of the Sun) and 0.5 days. For a number $n$ of observations, \citet{2002A&A...391..369K} indicated the number of statistically independent trial frequencies $n_f \sim n^{0.83}$. To recover short-period objects reported by \citet{2013MNRAS.429.2001H} (see below), we found that we had to take $n_f = 2 \times n^{0.83}$. This assumption shouldn't affect our ability to find putative transit signals. EEBLS also bins up the folded data into $n_\mathrm{b}$ bins, and for the search, we took $n_\mathrm{b} = 50$ to balance computational time against signal resolution (different from our chosen number of bins at other points in our analysis because the EEBLS search is more computationally intensive). In the limit that the orbital distance approaches the stellar surface, the fraction of an orbit spent in transit (called $q$ by EEBLS) is a maximum, 0.5. This value is the maximum ($q_\mathrm{max}$) we considered. The shortest transit to which \kepler~data would likely be sensitive is about 30 min, and so the minimum fraction of an orbit spent in transit we considered was 30 min/0.5 days = 0.042 ($q_\mathrm{min}$). 

\indent After applying EEBLS to these data, we had a collection of best-fit transit periods and depths (among other EEBLS statistics). We estimated the noise for each candidate light curve by folding the detrended data on this period, binning the points in 5-min wide bins, and taking the median of the errors for all bins as that light curve's noise. (Typical uncertainties were $\sim$ 10 parts-per-million ppm.) We dropped all targets with depths having signal-to-noise ratios SNR $< 3$ and signal detection efficiencies SDE $< 5$ \citep{2002A&A...391..369K}. These choices are motivated by eyeballing many transit candidates and are not necessarily optimized. 

\indent We also found that, for many targets, stellar oscillations mimicked transits but with unphysically long durations. To eliminate these signals as candidates, we applied an upper limit to transit durations $t_\mathrm{D}$. The ratio of the transit duration to the orbital period $P$ is given by
\begin{eqnarray}
\sin \left(\frac{\pi t_\mathrm{D}}{P}\right) &=& \left(\frac{R_\star}{a}\right) \sqrt{1 - b^2} \nonumber \\
&=& \left(\frac{3}{4 \pi}\right)^{1/3} \left( R_\odot/1\ \rm{AU} \right)\left( P/1\ \rm{yr} \right)^{-2/3} \left( \rho_\star/\rho_\odot \right)^{-1/3} \nonumber \\ &\times& \sqrt{1 - b^2}
\end{eqnarray}
where $\rho_\star/\rho_\odot$ is the stellar density normalized to the Sun's density and $b$ is the transit impact parameter. For a given $P$, the largest $t_\mathrm{D}/P$ in which we're interested corresponds to $b = 0$ (a central transit) and a minimum value for $\rho_\star/\rho_\odot$. We focused here on F, G, and K Sun-like stars, and so we take a minimum $\log(\rho_\star/\rho_\odot) = -0.3$ \citep{2000asqu.book.....C}:
\begin{eqnarray}
\max\bigg(\frac{\pi t_\mathrm{D}}{P}\bigg) = \sin^{-1}\left(\Gamma \times (P / 1\ \rm{yr})^{-2/3}\right)
\end{eqnarray}
where $\Gamma = \left(\frac{3\times10^{-0.3}}{4 \pi}\right)^{1/3}\times\left( R_\odot/1\ \rm{AU} \right) \approx 4 \times 10^{-3}$.

\indent After winnowing our list in these ways, we were still left with nearly 800 candidates. However, the signals for many of these candidates were still clearly not planetary transits, so we further selected out putative planets by eye, tossing signals that were apparently stellar oscillations or eclipsing/contact binaries -- typically they exhibited ellipsoidal variations (EVs) with amplitudes comparable to the occultation depths. This step involved a subjective assessment of each candidate that could potentially bias our survey. However, for this preliminary study, we were only concerned with the most obvious, easiest-to-analyze candidates. 

\indent Next, we removed any candidates that were reported as eclipsing binaries in \cite{2011AJ....142..160S}. We retained KIC 8435766 \citep{2013ApJ...774...54S} and 5080636 \citep{2013A&A...555A..58O}. We did not analyze the short-period candidates reported in \citet{2013MNRAS.429.2001H} but will in a future paper. Finally, we dropped all candidates that were clearly just discoveries reported in \citet{2013ApJS..204...24B} but with periods aliased to less than 0.5 days by EEBLS. For the remaining targets, we compared odd-even transits to check for consistency (Section \ref{sec:odd_even_transits}), which ensures that the right period is not twice the reported EEBLS period. We also checked even higher harmonics of the detected period to ensure detection of the correct period. We call this period the \emph{initial period}. In some cases, the initial period was larger than 0.5 days, and we removed these candidates for the purposes of this initial study. (However, our subsequent analysis showed that some candidates retained at this step actually ended up having final best-fit periods $>$ 0.5 days -- see Section \ref{sec:odd_even_transits}.) The MAST \kepler~archive lists some of our candidates as false positives, but we retain these candidates at this point -- our initial periods differ from the periods for those candidates given by the NASA Exoplanet Archive\footnote{\href{http://exoplanetarchive.ipac.caltech.edu}{http://exoplanetarchive.ipac.caltech.edu}}.

%

%

%

\begin{deluxetable*}{cccccccc}

\tablecaption{Stellar Parameters from the KIC and Quadratic Limb-Darkening Coefficients}

\tablehead{\colhead{KIC} & \colhead{$K_\mathrm{p}$} & \colhead{$T_\mathrm{eff}$} & \colhead{$\log(g)$} & \colhead{[Fe/H]} & \colhead{$R_\mathrm{\star}$} & \colhead{$\gamma_\mathrm{1}$} & \colhead{$\gamma_\mathrm{2}$} \\ 
\colhead{} & \colhead{} & \colhead{(K)} & \colhead{(cm/s$^2$)} & \colhead{} & \colhead{($R_\odot$)} & \colhead{} & \colhead{}} 
\startdata
2857722 & 15.012 & 6103 & 4.444 & -0.109 & 1.044 & 0.327778 & 0.295262 \\
2859299 & 13.791 & 5858 & 4.364 & -0.097 & 1.139 & 0.358787 & 0.281289 \\
3233043 & 13.787 & 4874 & 4.12 & -0.117 & 1.479 & 0.560457 & 0.165257 \\
4175105 & 14.538 & 5706 & 4.366 & -0.283 & 1.128 & 0.373013 & 0.271575 \\
4861364 & 13.404 & 5715 & 4.489 & -0.559 & 0.971 & 0.361508 & 0.272025 \\
5017876 & 13.713 & 5846 & 4.633 & -0.017 & 0.818 & 0.377622 & 0.274467 \\
5080636 & 14.404 & 3673 & 4.51 & 0.365 & 0.631 & 0.869878 & -0.0846 \\
5440651 & 15.341 & 5011 & 4.491 & -0.438 & 0.912 & 0.500786 & 0.202041 \\
5475494 & 15.145 & 6198 & 4.775 & -0.074 & 0.698 & 0.323981 & 0.298922 \\
5636648 & 15.33 & 5555 & 4.581 & -0.179 & 0.858 & 0.401317 & 0.262695 \\
5896439 & 13.863 & 6115 & 4.43 & -0.23 & 1.063 & 0.325149 & 0.295003 \\
6665064 & 15.183 & 5780 & 4.33 & -0.18 & 1.183 & 0.366403 & 0.276088 \\
7051984 & 12.771 & 5438 & 3.856 & 0.031 & 2.168 & 0.458598 & 0.229609 \\
7269881 & 14.409 & 4937 & 4.581 & -0.169 & 0.799 & 0.54707 & 0.172551 \\
7516809 & 15.525 & 5597 & 4.96 & -0.436 & 0.533 & 0.380383 & 0.271014 \\
7582691 & 15.174 & 3977 & 4.365 & -0.257 & 0.868 & 0.79709 & -0.00975153 \\
8260198 & 15.45 & 5176 & 4.545 & -0.569 & 0.865 & 0.44772 & 0.233505 \\
8435766 & 11.551 & 4957 & 4.454 & -0.093 & 0.948 & 0.547555 & 0.174685 \\
8588377 & 14.997 & 5514 & 4.329 & -0.384 & 1.169 & 0.395561 & 0.2597 \\
8645191 & 15.606 & 5445 & 4.638 & -0.14 & 0.789 & 0.426306 & 0.249731 \\
8703491\tablenotemark{a} & 12.19 & 5780 & 4.44 & 0 & 1 & 0.377656 & 0.272457 \\
9520443 & 15.896 & 4916 & 4.624 & 0.002 & 0.751 & 0.607948 & 0.12972 \\
9597729 & 14.84 & 5737 & 4.593 & -0.185 & 0.855 & 0.372468 & 0.275347 \\
9752973 & 10.047 & 5865 & 4.017 & -0.63 & 1.776 & 0.354771 & 0.267351 \\
9883561 & 15.473 & 4758 & 4.585 & 0.147 & 0.764 & 0.614604 & 0.130419 \\
9943435 & 14.397 & 5541 & 4.61 & -0.322 & 0.826 & 0.393983 & 0.264847 \\
10402660 & 15.769 & 4941 & 4.441 & -0.121 & 0.961 & 0.546715 & 0.174174 \\
10453521 & 14.208 & 6541 & 4.294 & -0.556 & 1.279 & 0.322989 & 0.275554 \\
11496490 & 13.866 & 6052 & 4.305 & -0.128 & 1.236 & 0.333308 & 0.291816 \\
11709423 & 15.503 & 5187 & 4.532 & -0.373 & 0.88 & 0.46458 & 0.22474 \\
11969092 & 15.23 & 4904 & 4.778 & 0.503 & 0.613 & 0.605463 & 0.142705 \\
11972387 & 12.435 & 6782 & 4.196 & 0.027 & 1.465 & 0.381328 & 0.236756 \\
12023078 & 12.65 & 5201 & 3.646 & -0.248 & 2.853 & 0.472991 & 0.215089 \\
12120286 & 14.633 & 5956 & 4.512 & -0.229 & 0.955 & 0.342166 & 0.287334 \\

\enddata
\tablecomments{The column labeled KIC shows the KIC number, $K_\mathrm{p}$ the stellar magnitude in the \kepler~bandpass, $T_\mathrm{eff}$ the stellar effective temperature, $\log(g)$ the base-10 log of the stellar surface gravity (in cm/s$^2$), [Fe/H] the stellar metallicity, $R_\mathrm{\star}$ the stellar radius (in solar units), and $\gamma_\mathrm{1/2}$ the quadratic limb-darkening coefficients.}
 \label{tbl:stellar_params}
\tablenotetext{a}{We assumed these parameters for KIC 8703491 were the same as the Sun's.}

\end{deluxetable*}%

\subsection{Odd-Even Transit Analysis}
\label{sec:odd_even_transits}
\indent To begin refining the candidate orbital periods, we first checked whether the correct periods for our candidates might actually be twice their initial periods. We detrended the data as above, masking out the transits, folded the data on twice the initial period, and binned these data into \overallnumbins bins.

\indent We fit transit light curves using the formalism from \citet{2002ApJ...580L.171M} and updated in \citet{2013PASP..125...83E}\footnote{\href{http://astroutils.astronomy.ohio-state.edu/exofast/}{http://astroutils.astronomy.ohio-state.edu/exofast/}}. We determined quadratic limb-darkening coefficients (LDCs), $\gamma_\mathrm{1/2}$, via tri-linear interpolation among the tables from \citet{2011A&A...529A..75C} using the stellar effective temperatures $T_\mathrm{eff}$, surface gravities $\log(g)$, and metallicities [Fe/H] from the \kepler~Input Catalog KIC\footnote{\href{http://archive.stsci.edu/kepler/kic.html}{http://archive.stsci.edu/kepler/kic.html}} \citep{2011AJ....142..112B} (shown in Table \ref{tbl:stellar_params}). The KIC does not contain these parameters for KIC 8703491, so we assumed solar values for that target (the photocenter analysis below shows it's probably a blended eclipsing binary anyway). The KIC does not provide uncertainties for individual stars, so we did not include any uncertainties in these stellar parameters in our analysis. As a result, the uncertainties on parameters related to the stellar properties (e.g. the impact parameter) are probably underestimated. More recent estimates of some stellar parameters are available (e.g., \citealt{2012ApJS..199...30P}) that suggest the KIC stellar parameters are biased for some stars. However, these estimates are not available for all our candidates, and so, rather than introducing non-uniform biases by using different stellar catalogs for different candidates, we elect just to use the parameters provided by the KIC.

\begin{figure}[h!]
\includegraphics[width=0.48\textwidth]{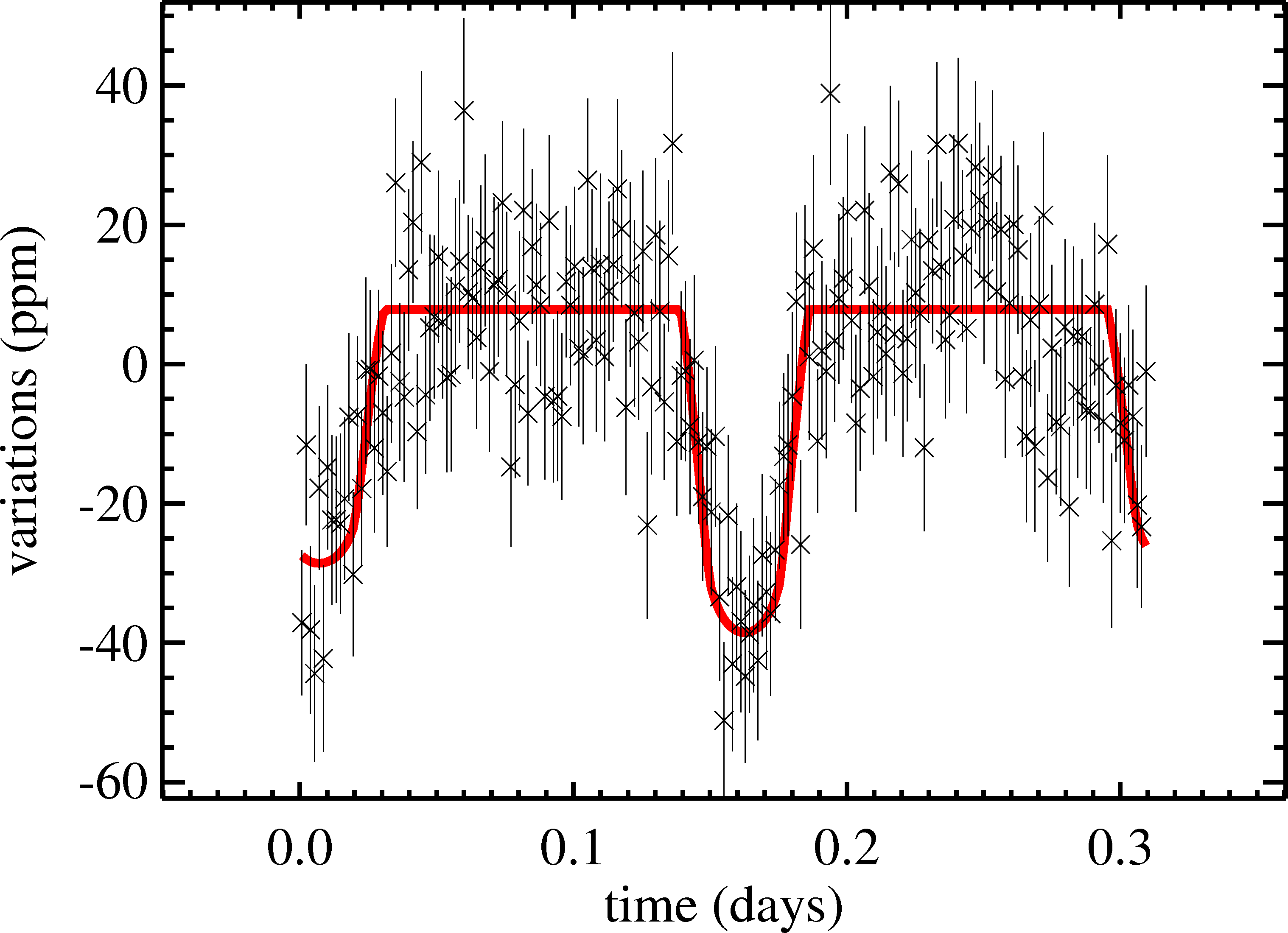}
\caption{Fit to the odd and even transits (red curve) for KIC 4861364, shown in parts-per-million ppm. The transit radii marginally disagree.}
\label{fig:KIC4861364_check_odd_even_transits}
\end{figure}

\indent Holding the LDCs fixed, we fit separate transit curves to the odd and even transits -- Figure \ref{fig:KIC4861364_check_odd_even_transits} shows one example. We fit for the orbital semi-major axis $a$ and impact parameter $b$ (both scaled to the sum of the objects' radii). The $a$-values were required to be greater than 1 and less than 10, and $b$-values to lie between 0 and 1. We fit mid-transit times for the odd and even transits $t_\mathrm{1/2}$ (requiring them to lie within half an orbital period of an initial eyeball estimate) and radius ratios for each transit $p_\mathrm{1/2}$ (required to lie between 0 and 1). We also fit for a shared out-of-transit baseline. Our model assumed circular orbits since we saw no clear indications of non-zero eccentricity, which tides likely would damp out for such short periods anyway. The model includes convolution of the transit signal over \kepler's 30-min integration times, which is particularly important for the short transits we consider here.

\indent We used a Levenberg-Marquardt (LM) $\chi^2$-minimization scheme\footnote{\href{http://www.physics.wisc.edu/~craigm/idl/down/mpfitfun.pro}{http://www.physics.wisc.edu/$\sim$craigm/idl/down/mpfitfun.pro}} to perform an initial fit and calculated the resulting reduced $\chi^2$, $\chi_\mathrm{red}^2$ -- often this procedure produced a $\chi_\mathrm{red}^2 > 1$, so we rescaled the data point uncertainties by $\sqrt{\chi_\mathrm{red}^2}$. Then we applied an MCMC analysis to find the best-fit parameters and their uncertainties. Following the recipe in \cite{2005AJ....129.1706F}, we initialized five chains with parameter values chosen randomly to lie between the boundaries given above. For each link in the chain, we proposed to update one or two model parameters at a time, with the proposed parameter(s) chosen randomly. Every 100 links, we calculated the fraction of accepted proposals, the acceptance fraction, and required it lie between 0.25 and 0.55. When it drifted out of this range, we dropped those last 100 links and started again. Every 1000 links, we masked out the first 10\% of the chain as a burn-in phase and calculated the Gelman-Rubin statistic $\hat{R}^{1/2}$ \citep{Gelman1992} to check the chains for convergence. Once the chains contained at least 5000 links and $\hat{R}^{1/2}$ for all chains dropped below 1.2, we considered the chains to have converged. The number of links was typically several thousand. For each parameter, we took the mean of all chains (with the burn-in phase dropped) as the best-fit parameter value. The distribution of values from the MCMC chains are not all distributed normally about the mean value for each parameter, so for the uncertainties, we took the upper/lower value of the distribution for which 34.1\% of the chain lay between the mean and the upper/lower value.

\indent We then compared the absolute difference between $p_\mathrm{1}$ and $p_\mathrm{2}$ normalized to their joint uncertainties added in quadrature, $|\Delta p|/\sigma$. If $|\Delta p|/\sigma > 3$ for a candidate, we chose twice the initial period for that system. If $|\Delta p|/\sigma < 1$, we chose the initial period. However, some cases had $1 \le |\Delta p|/\sigma \le 3$, so the correct period is ambiguous based on this comparison. For those candidates, we continued our analysis using both the initial and twice the initial period. Table \ref{tbl:check_odd_even_transits} shows the results of this comparison. 

\indent For those candidates with $|\Delta p|/\sigma > 3$, one of the transits may actually be a planetary eclipse, but only if its depth is consistent with a planet nearly in radiative equilibrium with its host star. In fact, \citet{2013ApJS..204...24B} used the depth of the apparent planetary eclipse to distinguish between planetary candidates and blended eclipsing binaries. However, we did not apply this criterion at this point to distinguish between planetary candidates and binaries. Instead, for those candidates with $|\Delta p|/\sigma > 3$, we allowed that the deeper of the two signals was the transit and the other was an eclipse. 

\indent In some cases, a best-fit period turned out to be $>$ 0.5 days at this point; we continued our analysis of those candidates. Also, in most cases, the $b$-values were poorly constrained by the data due to the severe convolution of the in/egress phases. However, this result does not significantly affect the $p$-value comparison.

\indent As an example of our procedure, consider the KIC 4861364 system. EEBLS reported an initial period of 0.155818 days. Comparison of the odd and even transit radii (Figure \ref{fig:KIC4861364_check_odd_even_transits}) shows marginal disagreement between transit radii at 1.58-$\sigma$, so we carried 0.155818 and (2 $\times$ 0.155818 =) 0.311636 days forward in our analysis. 

\indent As exemplified in Figure \ref{fig:KIC4861364_check_odd_even_transits}, several of the candidates exhibit EVs. Not including an EV model (as we do later in Section \ref{sec:fit_evs}) distorts slightly the transit fits. However, the fit for each transit is distorted in the same way (since EVs are symmetric), so the comparison still allows determination of the correct period.

\begin{figure}
\includegraphics[width=0.48\textwidth]{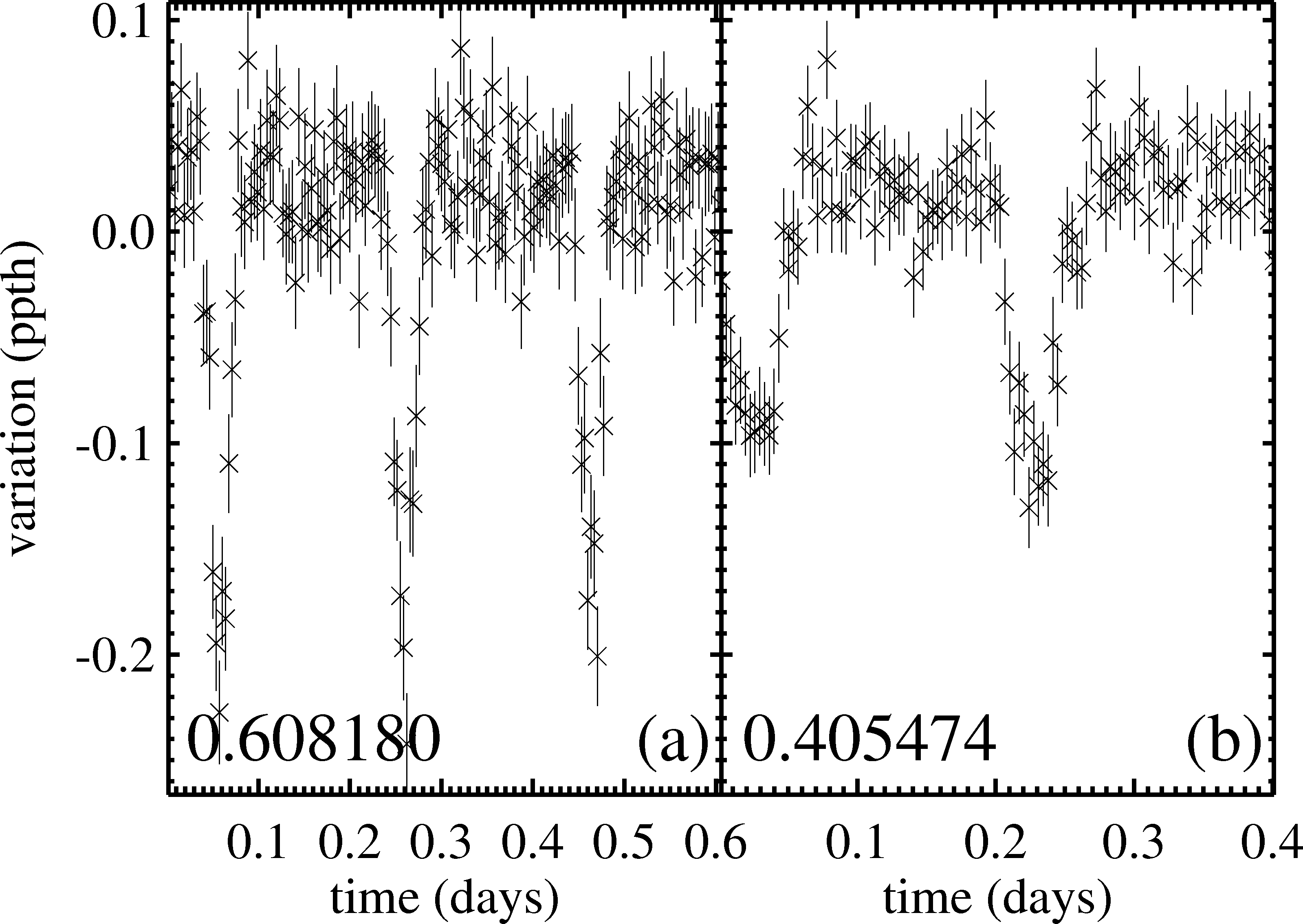}
\caption{Detrended, folded, and binned data for KIC 4175105. Panel (a) uses the period given by the NASA Exoplanet Archive, 0.60818 days, while (b) uses our twice our initial period ($2 \times$ 0.202737 =) 0.405474 days.}
\label{fig:KIC4175105_compare_exoplanet_archive}
\end{figure}

\indent Several of our candidates are listed as false positives on the MAST \kepler~archive but with periods given on the NASA Exoplanet Archive different from our initial periods. As shown in Section \ref{sec:photocenter_analysis}, we confirmed that these candidates are likely false positives on the basis of in-transit photocenter shifts, using our initial periods. Interestingly, in some cases, the Archive period for a candidate is nearly an integer multiple of our initial period. Figure \ref{fig:KIC4175105_compare_exoplanet_archive} shows an example. Panel (a) shows the detrended data, folded and binned on the Archive period (0.60818 days), while (b) shows the same data, folded and binned on twice our initial period ($2 \times 0.202737 =$) 0.405474 days as part of our odd-even transit comparison. (The apparent difference in transit depth and duration arises because our initial period is slightly incorrect, and so the folding and binning distorts the transit.)

\indent The Archive's period is almost exactly three times our initial period, and the figure clearly shows multiple transits for both periods, suggesting our initial period (0.202737 days) is closer to the correct one. Similar results apply for the following candidates: 4175105, 8645191, 9597729, 9752973, 11496490, 11969092, and 12023078. Based on the \kepler~data validation reports\footnote{The data validation reports provide details of the \kepler~mission's assessment of the false positive status of a candidate (among other things) and are available here: \href{http://exoplanetarchive.ipac.caltech.edu/cgi-bin/ExoTables/nph-exotbls?dataset=cumulative}{http://exoplanetarchive.ipac.caltech.edu/cgi-bin/ExoTables/nph-exotbls?dataset=cumulative}. As of 2013 Oct 2, to access the report for a candidate, click on the ``i'' icon next to the candidate's KIC number in the table, and then select ``Kepler Pipeline KOI Overview'' from the pop-up menu. A link to the corresponding report appears near the bottom of the page that opens.} for these candidates, it seems the determination of the orbital periods involved folded and binning the data into bins 30-min wide. The resulting transits often comprised a single data point, likely making it difficult for the period-determination scheme to spot the multiple transits and recognize that the wrong period had been chosen. Choosing the optimal time resolution for folded and binned data is an important issue in a search for transits of very short-period planets.

\subsection{Photocenter Analysis} 
\label{sec:photocenter_analysis}

\indent To winnow our candidate list further, we next looked for statistically significant in-transit photocenter shifts. Since \kepler's pixels span 3.98 arcsecs on one side, flux is often blended between \kepler~targets and other objects nearby in the sky. This blending dilutes transit signals and can, for example, make the occultation of an eclipsing binary look like a planetary transit. \citet{2010ApJ...713L.103B} suggested small shifts in the photocenters of \kepler~target stars during transit can indicate blending of light with nearby objects. Since the transiting object blocks some of the target star's contribution, the flux-weighted photocenter may shift toward a nearby untransited object. Fortunately, the \kepler~team provides photocenter data (via the `MOM\_CENTR*' and `POS\_CORR*' fields in the light curve FITS files). 

\indent For each of our candidates, we collected the raw flux-weighted photocenter x/y positions (`MOM\_CENTR*') and subtracted the reference pixel value (the `CRVAL*P' keyword in the `APERTURE' extension of the light curve FITS files) and then the position correction values (`POS\_CORR*'). These data still exhibited systematic trends from quarter to quarter, so we applied the same mean boxcar filtering that we applied to the flux data. We then converted these pixel positions to sky positions, right ascension (R.A.) and declination (decl.), using the IDL routine xyad.pro\footnote{\href{http://idlastro.gsfc.nasa.gov/ftp/pro/astrom/xyad.pro}{http://idlastro.gsfc.nasa.gov/ftp/pro/astrom/xyad.pro}} -- Figure 6 shows an example dataset. For each RA/Dec datum in each quarter, we calculated the offset from the mean RA/Dec-value for that quarter. Collecting together (but not folding and binning) these data from all quarters, we calculated the mean and $\sigma$ for the in-transit and for the out-of-transit RA/Dec positions. (Note that we excluded points spanning the possible eclipse for a candidate from the collection of out-of-transit points since some candidates exhibited deep eclipses -- in many cases, probably too deep for the candidate to be a planet -- that would skew the out-of-transit photocenter position.) To estimate the uncertainties on each unfolded/unbinned photocenter data point in and out of transit, we used the LM algorithm to fit a 0th-order polynomial and took the $\sqrt{\chi_\mathrm{red}^2}$ for the in-/out-of-transit fits as the respective uncertainties. Then we re-fit the 0th-order polynomial using the MCMC analysis to estimate the best-fit RA/Dec positions and (possibly) asymmetric uncertainties for the in- and out-of-transit data separately. We then divided the difference between the mean in- and out-of-transit RA/Dec positions by their joint $\sigma$. Table \ref{tbl:all_binned_photocenter_table} shows this difference. For those candidates whose periods were still ambiguous, we conducted this analysis for both periods. In all cases for which a significant shift was detected at one trial period, it was also detected at the other. 

\indent Of our analyzed candidates, all but \numnophotoshift showed $\ge$ 3-$\sigma$ in-transit photocenter offsets in either RA or Dec. For simplicity, we discontinued analysis of objects showing $>$ 3-$\sigma$ photocenter offsets at this point. It is possible that the transits of actual planets can cause small photocenter shifts, given the right proximity and brightness for blended objects, but follow-up, high spatial resolution observations are usually required to sort those cases out.

\begin{figure}
\includegraphics[width=0.48\textwidth]{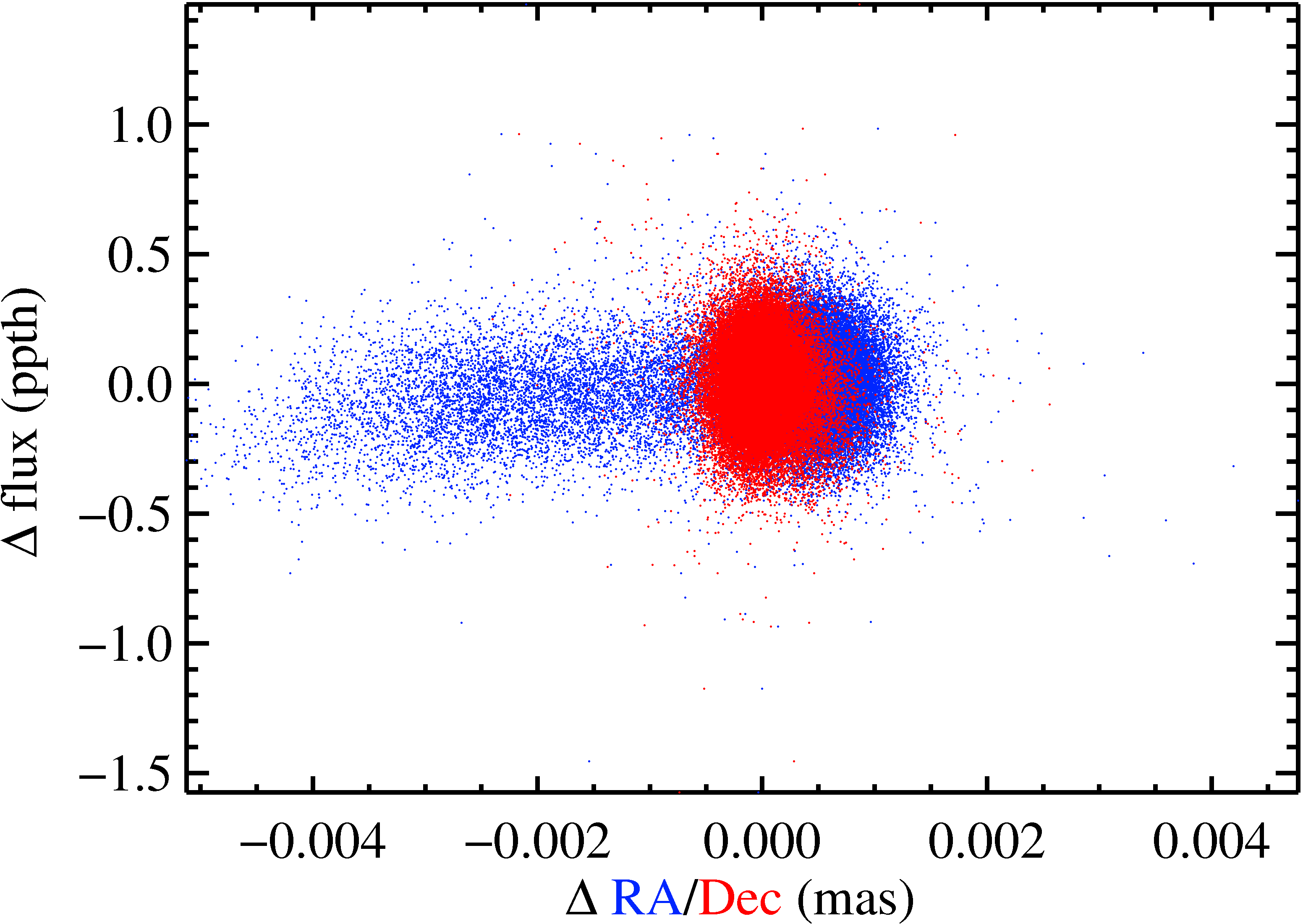}
\caption{Flux variations ($\Delta$ flux in parts-per-thousand ppth) vs. photocenter variations (in milliarcsecs mas) for KIC 12023078. Blue/red points indicate RA/Dec. This object shows clear photocenter variation correlated with the flux variations, suggesting it suffers blending and dilution.}
\label{fig:KIC12023078_binned_photocenter}
\end{figure}

\subsection{Refining the Orbital Periods and Looking for Transit-Timing Variations}
\label{sec:refining_periods}

\indent For those candidates with $<$ 3-$\sigma$ in-transit photocenter shifts, we next refined the orbital periods, estimated their uncertainties, and looked for transit-timing variations (TTVs). We generated a template transit by using the LM algorithm to fit a transit light curve to all of a candidate's data, folded and binned on the candidate's revised best-fit period (either initial or twice the initial period). We then attempted to fit that template transit to the data not folded/binned on the orbital period, holding fixed all the transit parameters except the mid-transit time and the out-of-transit baseline. As an initial estimate for the uncertainty of each unfolded/unbinned data point, we took the mean of $\sqrt{N} \sigma$ from all the \emph{binned} data for a candidate, where $N$ is the number of data points in a bin and $\sigma$ the uncertainty estimated for that bin.

\indent For those candidates with transit depths three times larger than this estimated uncertainty (SNR $>$ 3), we analyzed the individual transits. Considering each section of unfolded/unbinned data that spanned a full orbit, we fit the individual transit mid-times $t_\mathrm{n}$ using the LM routine and the template transit but without assuming a linear ephemeris. We required out-of-transit flux variations to be less than the transit depth, and if they were not for a particular span of data, we did not analyze that span. The LM routine provided estimates of parameter uncertainties (via the covariance matrix), which we re-scaled by $\sqrt{\chi_\mathrm{red}^2}$. As long as at least six points remained, we removed data with uncertainties exceeding 4-$\sigma$ for the overall distribution of uncertainties to exclude unusually poor fits. 

\indent Next, we used the LM routine to fit an ephemeris to these $t_\mathrm{n}$, allowing for secular drift in the orbital period by including a term quadratic in $n$. (Tidal decay of the orbit over the observational baseline could give such secular drift.) We removed $t_\mathrm{n}$-values more than 4-$\sigma$ from this fit and rescaled the associated uncertainties by the resulting $\sqrt{\chi_\mathrm{red}^2}$. Typical uncertainties were a few minutes. 

\begin{figure}
\includegraphics[width=0.48\textwidth]{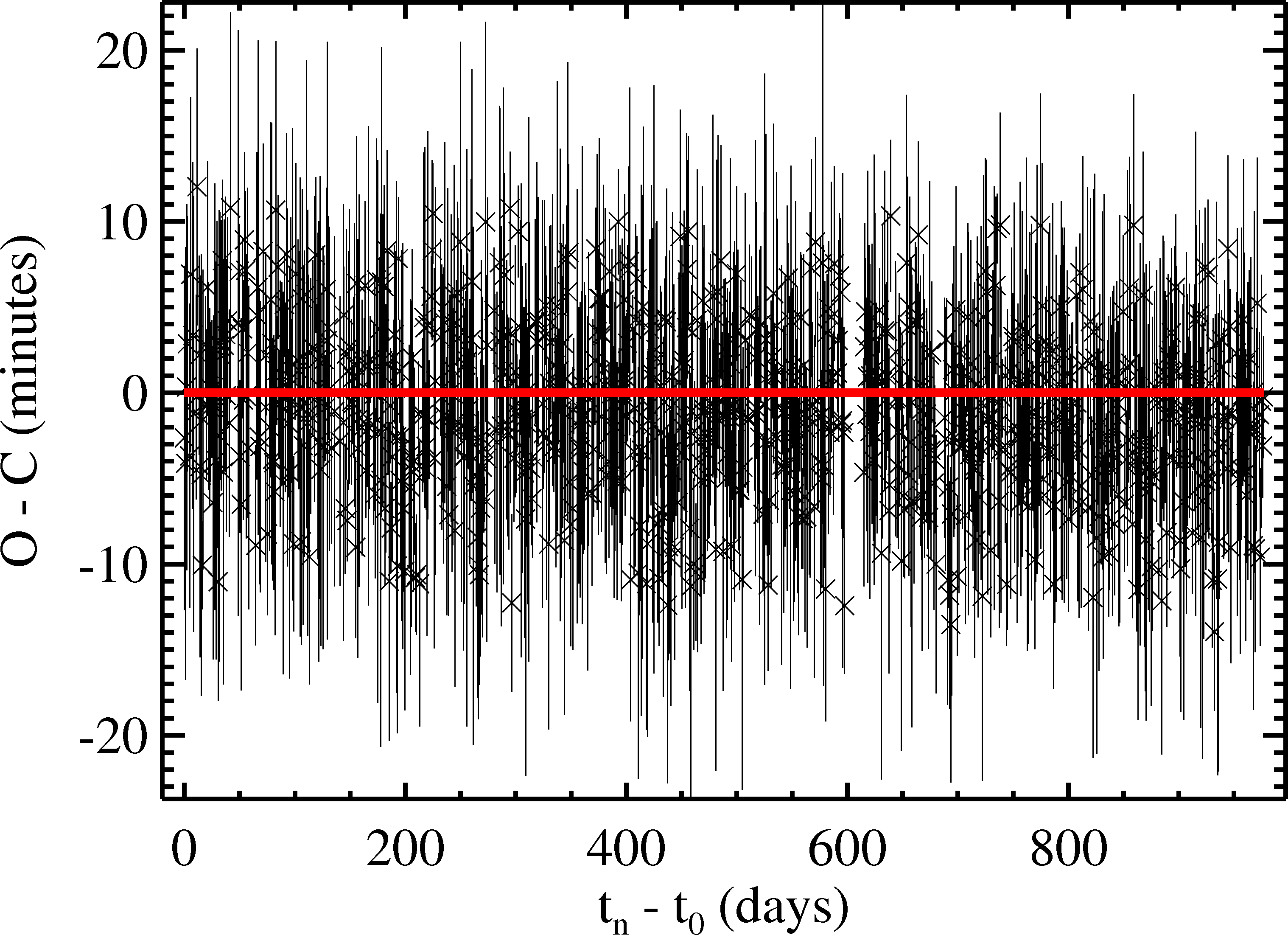}
\caption{Transit times $t_\mathrm{n}$ for KIC 8435766 (x-axis) with a linear ephemeris subtracted out $t_\mathrm{n} - n \times P$ (y-axis) to give the O - C-values. This system does not exhibit statistically robust (3-$\sigma$) TTVs, and the red line does not show a fit but is rather a flat line.}
\label{fig:KIC8435766_fit_each_mid}
\end{figure}

\indent Then, we applied the MCMC analysis to fit the quadratic ephemeris to determine a refined period, period drift, and uncertainties. To search for periodic transit-timing variations (TTVs), we subtracted the best-fit linear portion of the ephemeris and calculated a Lomb-Scargle periodogram for the resulting values, which represent the difference between the observed and calculated (for a linear ephemeris) transit times (O - C). We sought signals with false alarm probabilities exceeding 1\%. Figure \ref{fig:KIC8435766_fit_each_mid} shows an example O - C plot for KIC 8437566, along with a flat line showing no secular TTVs. 

\indent For most of our candidates, however, individual transits are too shallow to detect. In these cases, we found that estimating the $t_\mathrm{n}$ as above amounted to fitting noise. Instead, we folded and binned several adjacent transits together on the initially estimated period. We determined the number of foldings by comparing the uncertainties on unfolded/unbinned data points and the full transit depth, giving an estimate of the transit SNR. We folded and binned together at least $N_\mathrm{orb} = 3/\mathrm{SNR}^2$ transits to ensure that the averaged transits were likely to be detected at 3-$\sigma$. Folding together $N_\mathrm{orb}$ adjacent transits and binning in 30-min wide bins, we determined the $t_\mathrm{n}$ for these averaged transits by applying nearly the same analysis as above, except that we shifted the $t_\mathrm{n}$ (which, because of the folding, lay between 0 and the orbital period) by the average of the minimum and maximum times spanned by the $N_\mathrm{orb}$ adjacent transits. Of course, with this approach, we will average out TTVs with periods shorter than $N_\mathrm{orb}$ times the period, but we couldn't detect those anyway since individual transits lay below the noise. We conducted numerical experiments with synthetic datasets (but with the same sampling as the real data) and found that we could robustly recover secular TTVs with this technique, as long as the total variation in O - C over the observational baseline exceeded the intrinsic scatter. 

\indent The best-fit periods, uncertainties, and mid-transit times for the first transits ($t_\mathrm{0}$) are shown in Table \ref{tbl:fit_evs}. None of our objects show statistically significant secular or periodic TTVs based on this analysis.

\subsection{Fitting a More Complete Photometric Model}
\label{sec:fit_evs}
\indent Finally, we fit a more complete photometric model to the data, which included ellipsoidal variations, the Doppler beaming signal, and a phase curve for reflected/emitted flux. As in \citet{2013ApJ...774...54S}, we modeled these signals as the sum of sinusoids:
\begin{equation}
\Delta F = A_\mathrm{DB} \sin(2\pi \phi) - A_\mathrm{ELV} \cos(2\times 2\pi \phi) - A_\mathrm{ill} \cos(2\pi \phi)
\end{equation}
where $A_\mathrm{DB}$ is the amplitude of the Doppler beaming signal, $A_\mathrm{ELV}$ the amplitude of the ellipsoidal variations, $A_\mathrm{ill}$ the amplitude of the phase curve, and $\phi$ the orbital phase ($= 0$ at mid-transit and 0.5 at mid-eclipse). Each of these amplitudes was a free parameter, required to be greater than 0. Our planet candidates might be close enough to their host stars ($a \rightarrow 1$) that higher-order corrections to the ellipsoidal variations model are necessary \citep{2012ApJ...751..112J}, but we leave that aspect for future work. We also included a trapezoidal eclipse centered on orbital phase $\phi = 0.5$, with the eclipse depth $D$ a free parameter (allowed to drift between 0 and 1) and not tied to $A_\mathrm{ill}$. We fit the transit using the same model as in Section \ref{sec:odd_even_transits}, assuming that transit signal with the largest $p$-value from the analysis in Section \ref{sec:odd_even_transits} was actually the transit and the dip (if any) half an orbit later was a planetary eclipse. We applied the same combination of LM algorithms, error bar rescaling, and MCMC analysis as in Section \ref{sec:odd_even_transits}.

\section{Results}
\label{sec:results}

Table \ref{tbl:derived_params}  shows the physical parameters for our planetary candidates derived from the photometric model and the stellar parameters given the KIC, and Figures 8-11 show the data and model fits.

\begin{figure}
\includegraphics[width=0.48\textwidth]{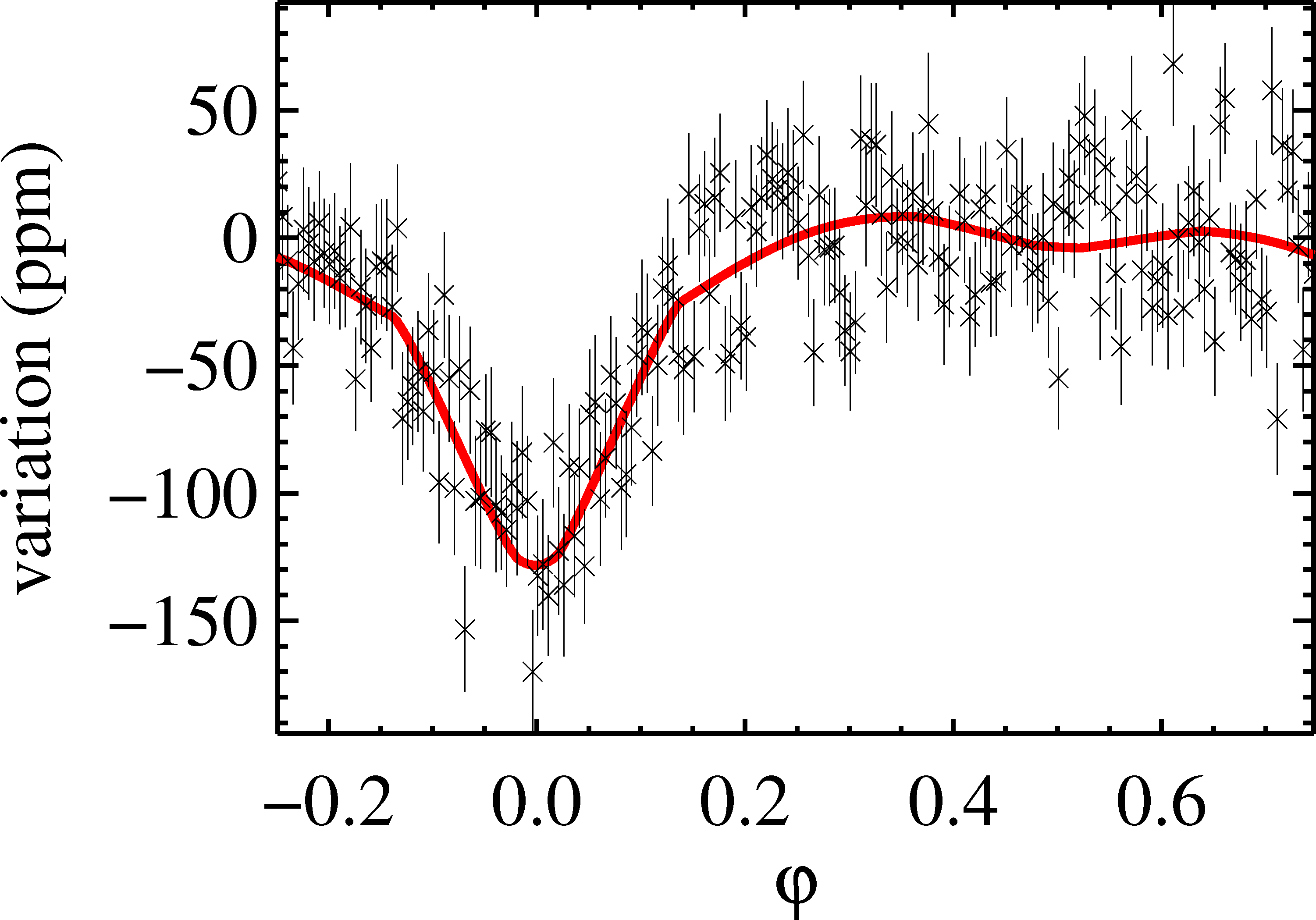}
\caption{Photometric model fit and data for KIC 5080636, with a period of 0.176896 days.}
\label{fig:KIC5080636_fit_EVs_MCMC}
\end{figure}

\begin{figure}
\includegraphics[width=0.48\textwidth]{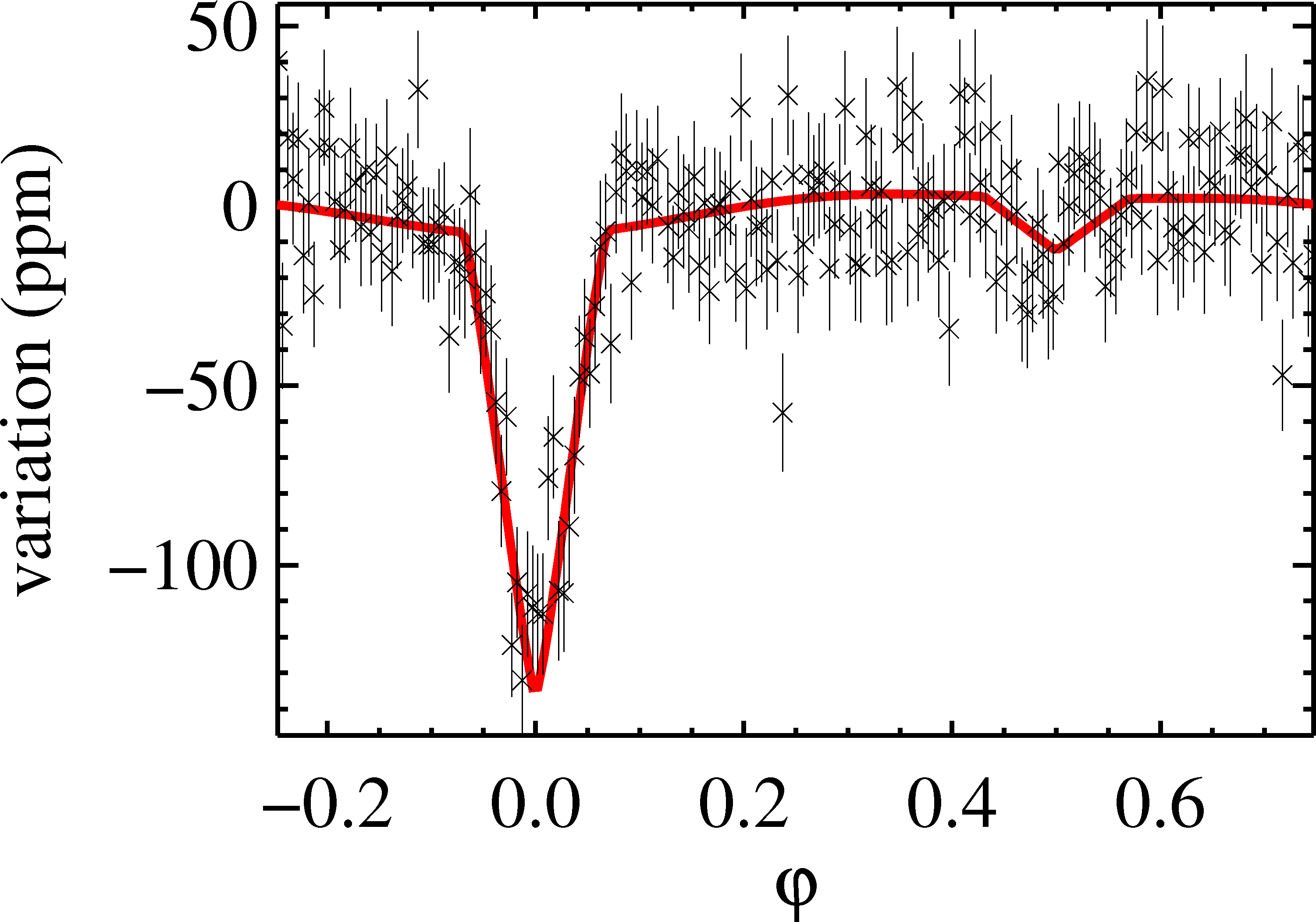}
\caption{Photometric model fit and data for KIC 7269881, with a period of 0.306936 days.}
\label{fig:KIC7269881_fit_EVs_MCMC}
\end{figure}

\begin{figure}
\includegraphics[width=0.48\textwidth]{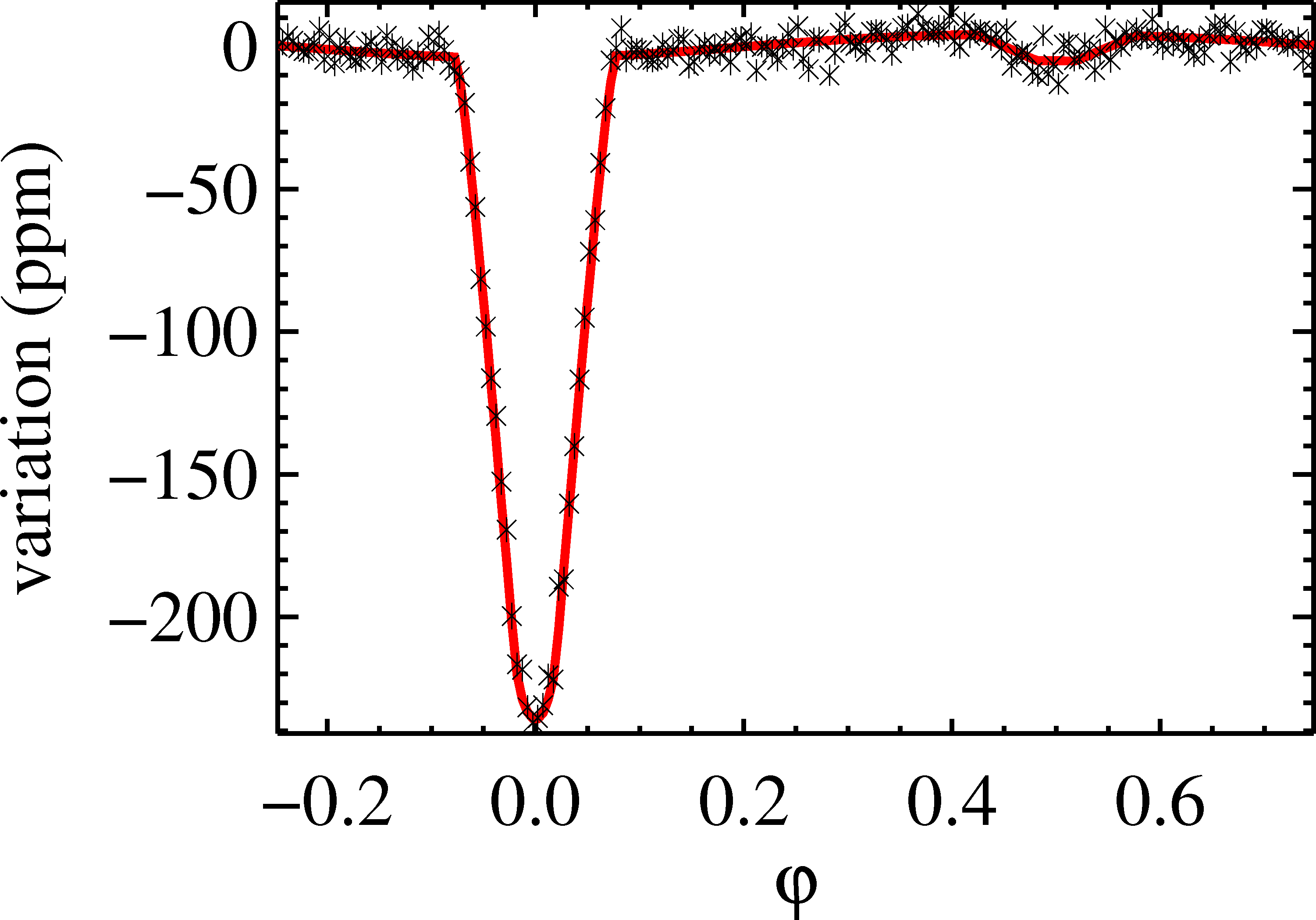}
\caption{Photometric model fit and data for KIC 8435766, with a period of 0.3550080 days.}
\label{fig:KIC8435766_fit_EVs_MCMC}
\end{figure}

\begin{figure}
\includegraphics[width=0.48\textwidth]{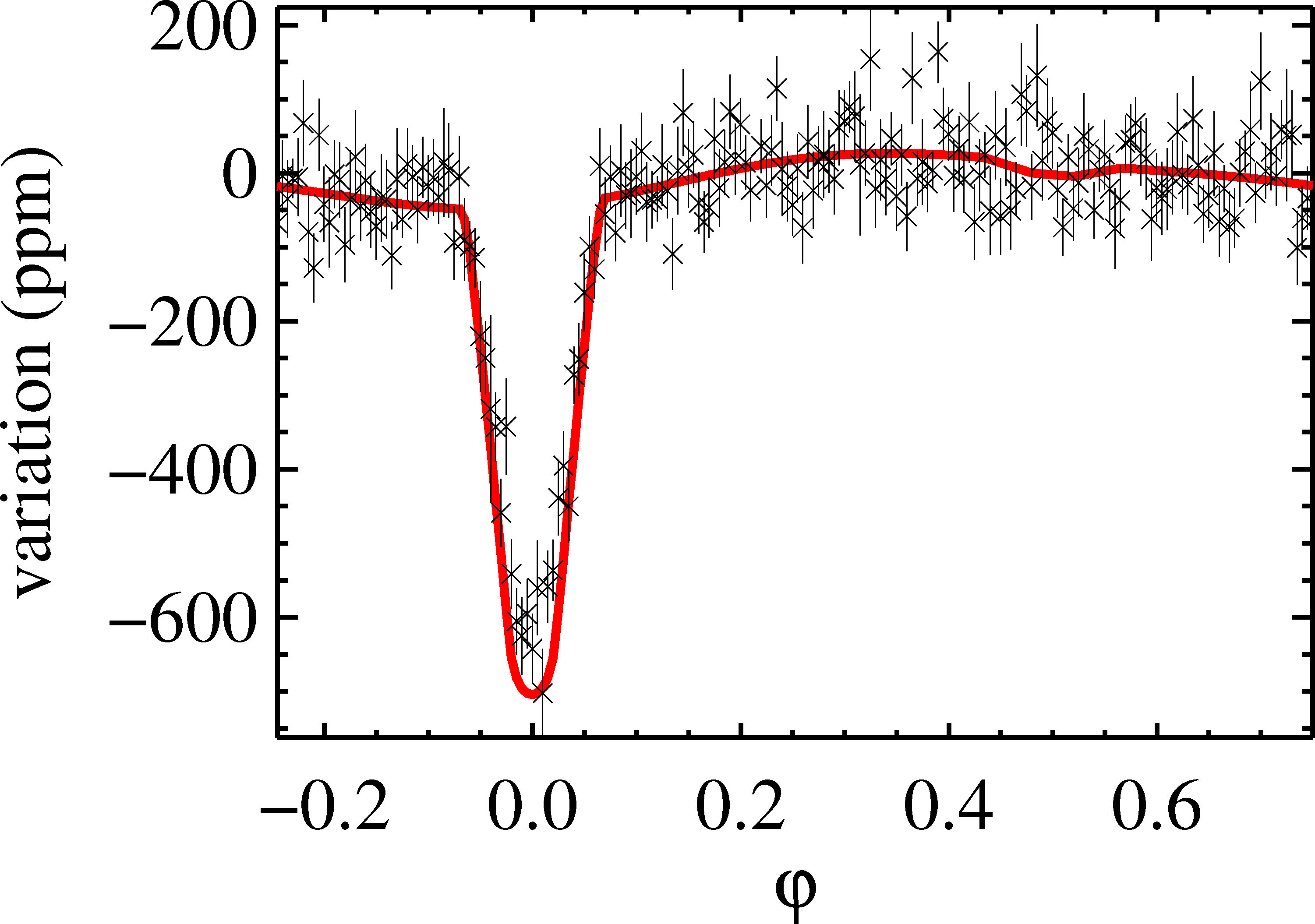}
\caption{Photometric model fit and data for KIC 10453521, with a period of 0.44841 days.}
\label{fig:KIC10453521_fit_EVs_MCMC}
\end{figure}

\indent Table \ref{tbl:derived_params}  shows the physical parameters for our planetary candidates derived from the photometric model and the stellar parameters given the KIC, and Figures \ref{fig:KIC5080636_fit_EVs_MCMC} - \ref{fig:KIC10453521_fit_EVs_MCMC} show the data and model fits. In the table, $u$ is the linear limb-darkening coefficient and $g$ the gravity-darkening exponent, both derived from interpolation among the tables from \citet{2011A&A...529A..75C}. $R_\mathrm{p}$ is the planetary radius (in Earth radii $R_\mathrm{Earth}$). $T_\mathrm{p, eq}$ is the blackbody radiative equilibrium temperature for a uniform dayside, based on the corresponding $T_\mathrm{eff}$ from the KIC and the best-fit $a$-value as $T_\mathrm{p, eq} = \left(\sqrt{2} a\right)^{-1/2} T_\mathrm{eff}$. $T_\mathrm{p, meas}$ is the dayside blackbody temperature estimated from the eclipse depth $D$ (and incorporating convolution of the blackbody radiation curve with \kepler's response function\footnote{\href{http://keplergo.arc.nasa.gov/CalibrationResponse.shtml}{http://keplergo.arc.nasa.gov/CalibrationResponse.shtml}}). Given the typically large temperature uncertainties, we chose not to incorporate the small geometric effects from the finite stellar size and finite distance between the planetary candidate and star. $q_\mathrm{DB}$ is the mass ratio determined from the Doppler beaming amplitude (see Equation 1 in \citealt{2011AJ....142..195S}), and $q_\mathrm{ELV}$ the mass ratio determined from the ellipsoidal variation amplitude (see Equation 7 in \citealt{2010A&A...521L..59M}). Using the KIC stellar parameters, we converted these mass ratios to the corresponding planetary masses $M_\mathrm{p, x}$. For most candidates, we did not robustly detect the Doppler beaming and ellipsoidal variation signals (at 3-$\sigma$, the amplitudes are consistent with zero), and so the upper limits on $q$- and $M_\mathrm{p, x}$-values for those candidates are best viewed as upper limits for a non-detection. 

\indent We discuss individual candidates below, but we do not decide whether a candidate is a planet. Such a determination generally requires follow-up observations, and we have such a campaign in the works. The model fit parameters for many of our candidates are consistent with a planetary interpretation, but for others, the parameters (usually the eclipse depth) make a stellar interpretation more plausible. Also, \citet{2013ApJ...774...54S} pointed out that, for a planet in radiative equilibrium with its host star and no transport of stellar heating, the night side contributes no flux, and so $D$ should be $\approx 2 A_\mathrm{ill}$. We discuss this point below for some candidates but don't use it to distinguish planets since, in many cases, we couldn't clearly detect an eclipse and the model fit often accounted for any systematic drifts in the data as a non-zero $A_\mathrm{ill}$.

\subsection{KIC 5080636}
\label{sec:KIC5080636}


\indent \citet{2013A&A...555A..58O} reported this candidate as part of the KOI-1843 system, which already had two planetary candidates (with periods 4.2 and 6.4 days). The fact that this candidate is a member of a system with two other candidates increases the probability that it is indeed a planet \citep{2012ApJ...750..112L}. Although the analysis in \citet{2013A&A...555A..58O} suffers from ``numerical problems'', the transit parameters from that study agree with ours to better than 3-$\sigma$. Our additional analysis of the eclipse depth gives an effective temperature consistent with that expected for radiative equilibrium. However, $D/A_\mathrm{ill} = 0.31 \pm 0.28$ is inconsistent with the planetary interpretation.

%

\indent \citet{2013ApJ...773L..15R} also analyzed the data for this object with a transit and photometric model similar to ours. Our transit and orbital parameters all agree with theirs to within 3-$\sigma$, but our photometric parameters are typically larger, particularly our $A_\mathrm{ill}$-value. As a check, we generated synthetic data with time sampling and uncertainties that mimicked our KIC 5080636 data, except we set $A_\mathrm{ill}$ to zero, and we found that we recovered the zero value. \citet{2013ApJ...773L..15R} analyzed short-cadence data (with a time resolution of 1-min) and apparently did not include the possibility of a Doppler signal (we did not detect a robust Doppler signal). These differences might account for the discrepant photometric parameters. 


\subsection{KIC 7269881}
\label{sec:KIC7269881}

\indent Model parameters for this candidate are also consistent with a planetary interpretation, although $T_\mathrm{p, meas}$ is larger than $T_\mathrm{p, eq}$ by about 2.4-$\sigma$.

\subsection{KIC 8435766}
\label{sec:KIC8435766}

\indent All our model parameters are consistent with those reported in \citet{2013ApJ...774...54S}, where this candidate's discovery was first announced, and bolster the planetary interpretation.\footnote{In fact, \citet{Howard2013} and \citet{Pepe2013} have confirmed the planetary nature of this object via RV observations.}


\subsection{KIC 10453521}
\label{sec:KIC10453521}

\indent \kepler~only returned data from Q10 for this system, which is why its $t_\mathrm{0}$-value is so large. The $T_\mathrm{p, meas}$ and $T_\mathrm{p, eq}$-values are consistent with a planetary interpretation for this system. However, the eclipse is only detected at 2-$\sigma$, while we detected the DB and $A_\mathrm{ill}$ signals at 3.6-$\sigma$ and 5-$\sigma$ respectively. The corresponding DB mass is 5000 $M_\mathrm{Earth}$, and the $A_\mathrm{ill}$ is larger than the eclipse depth. It's plausible that this system is a blended eclipsing binary but with a period twice our reported period (0.89 days) and that the $A_\mathrm{ill}$ signal is actually an ellipsoidal variation signal. The large impact parameter also points to the binary star interpretation. The apparent radius (3.8 $R_\mathrm{Earth}$) is much larger than expected for a rocky planet but may be consistent with a super-Earth having a hydrogen-rich atmosphere \citep{2012ApJ...761...59L}. For a Neptune-like density, such a planet might not be disrupted at this orbital distance.

\section{Discussion and Conclusions}
\label{sec:discussion_and_conclusions}

\indent Our search for very short-period planets in the \kepler~dataset has turned up \numnophotoshift candidates with best-fit periods $<$ 0.5 days, two of which were reported elsewhere (KIC 5080636 and 8435766). Although our search is unique and complementary to other \kepler~searches, our selection criteria for which systems' datasets to analyze involved subjective assessments, and so we likely introduced important but unquantified biases into the survey. Future work will involve careful study of the biases for such a survey to determine the frequency of such planets. However, the fact that we only found a handful suggests such planets are rare.

\indent Confirming whether our candidates are planets, though, will require additional follow-up observations and analysis. Given their proximity to their host stars, our planetary candidates may be well suited among \kepler~rocky planet candidates for RV follow-up. For example, a 10-$M_\mathrm{Earth}$ planet with an orbital period of 5 hours induces a radial velocity in a solar-mass star of 10 m/s (and the mass-radius relations from \citealt{2007ApJ...659.1661F} suggest such a massive planet can have a radius comparable to those we found). Sufficient RV precision to observe such a signal is currently achievable from ground-based instruments (e.g. \citealp{2011ApJ...729...27B}). On the other hand, three of our four candidates have \kepler~magnitudes greater than 14, making RV follow-up very difficult.

\indent In addition to many other planet types, the upcoming \tess~mission\footnote{\href{http://tinyurl.com/kc836c7
}{http://tinyurl.com/kc836c7}} will be ideally suited for finding more planets with very short-periods. In particular, its observational cadence length of 1-min will allow the mission to spot short-duration/small-depth transits that are convoluted beyond detection by \kepler's 30-min cadence. This possibility may be counterbalanced by the reduced sensitivity of \tess. In any case, such planets could represent a large fraction of \tess~discoveries and would be among the most amenable to detailed follow-up.

\indent These very short-period planets may also serve as sensitive probes for planet-star interactions and stellar magnetospheres. Several studies give tantalizing evidence for planet-stellar magnetosphere interactions \citep{2010EGUGA..1213591S}. With surface temperatures $>$ 2000 K, these short-period planets may shed rock vapor atmospheres and lose significant mass \citep{2013MNRAS.433.2294P, 2010MNRAS.407..910J}, and the interaction between the atmospheres and stellar wind may be observable, as for Mercury \citep{2008merc.book..251K}. Although such detection would probably require a significant investment of observational facilities, detection of such an atmosphere could provide direct constraints on the composition of a rocky exoplanet. 

\indent In fact, these rock vapor atmospheres may be sufficiently dense (possibly surface pressures of millibars -- \citealp{2009ApJ...703L.113S}) that they effect day-night redistribution of stellar heating, or internal transport of stellar heating (possibly via degree-1 convection -- \citealp{2011ApJ...735...72G}) may significantly warm a planet's night side. Either process might measurably modify the planetary phase curve from the sinusoidal one assumed here.

\indent Whatever the prospects for follow-up of very short-period planets, their origins are puzzling, and it's not clear that the usual origin scenarios for close-in planets apply. 

\indent So close to their host stars, very short-period planets probably didn't form where we see them now since temperatures in the protoplanetary gas disk probably exceeded dust condensation temperatures \citep{2013MNRAS.431.3444C}, and these host stars were likely distended during the pre-main sequence, with radii several times larger than at present, preventing planet formation at such distances \citep{1994ApJS...90..467D}.

\indent Gas disk migration of solid planets could potentially bring the planets in from farther out \citep{2000prpl.conf.1111L} but would require the planets to have migrated much closer to their host stars than typical hot Jupiters. Moreover, interactions with the young star's magnetosphere may have truncated the gas disk at a distance of several stellar radii, where the gas orbited with a period of several days, and inward migration of the planet probably should have stalled once it crossed the 2:1 mean motion resonance with gas disk's inner edge \citep{1996Natur.380..606L}. 

\indent It is possible that the inward migration of a gas giant planet could capture what becomes a very short-period planet into resonance and then the latter planet could migrate in, along with the more massive planet \citep{2007ApJ...660..823M}. In this case, very short-period planets should be accompanied by more massive, outer planets. Our very short-period candidate in the KIC 5080636 system is accompanied by two longer period planetary candidates (about 4 and 6 days), but their radii (1.35 and 0.84 $R_\mathrm{Earth}$, respectively -- \citealp{2013ApJS..204...24B}) suggest they are probably not gas giants. We will search for longer period companions in our candidate planetary systems in future work.

\begin{figure}
\includegraphics[width=0.48\textwidth]{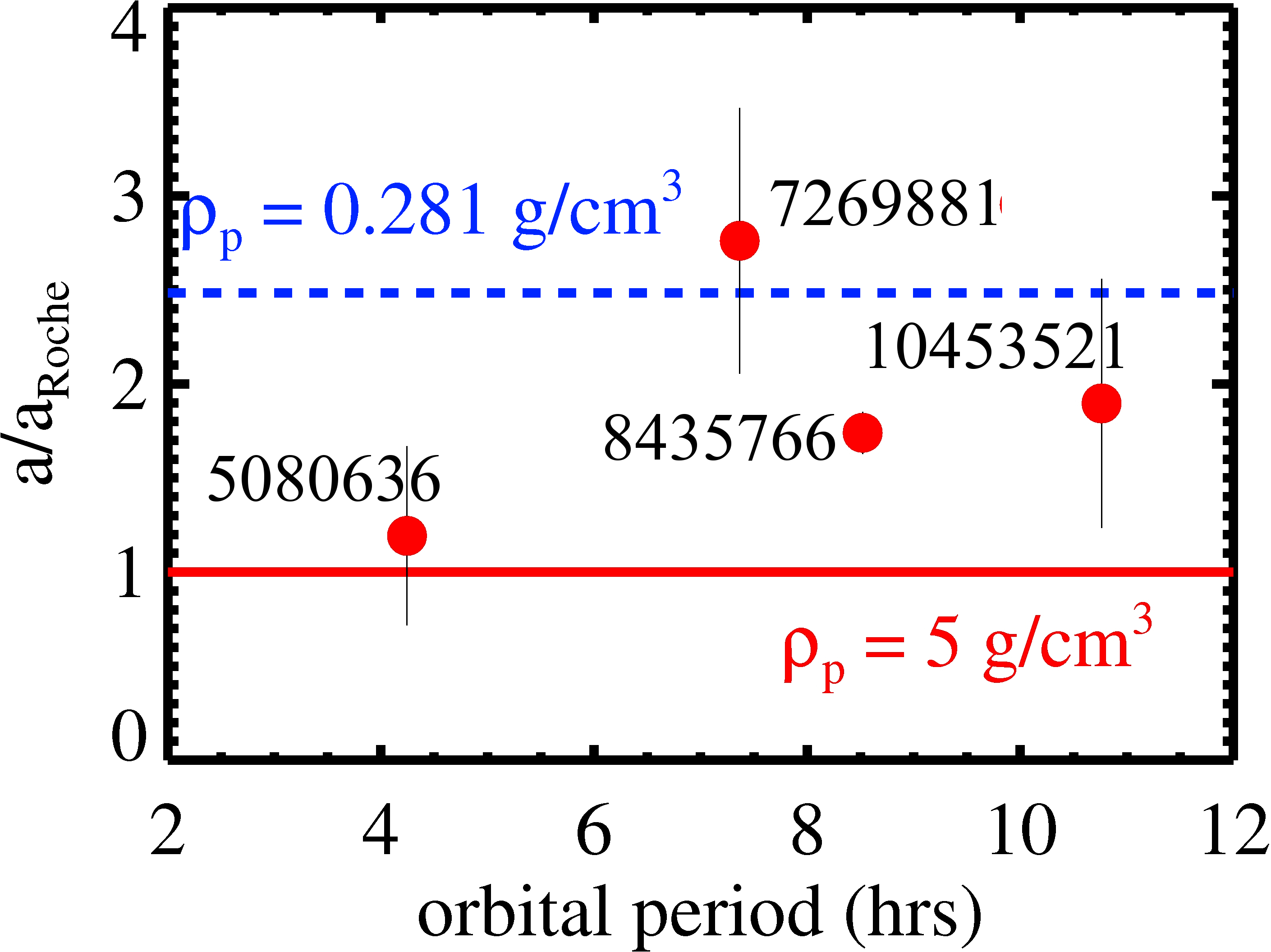}
\caption{The ratio of the semi-major axis $a$ to the Roche limit $a_\mathrm{Roche}$. The red circles show this ratio for each of our planetary candidates, assuming an Earth-like density of $\rho_\mathrm{p} = 5$ g/cm$^3$. The solid red line shows where $a/a_\mathrm{Roche} = 1$ for that density, while the dashed blue line shows where $a/a_\mathrm{Roche} = 1$ if all the points were shifted for a WASP-12 b-like density of 0.281 g/cm$^3$.}
\label{fig:roche_lobe}
\end{figure}

\indent Planet-planet scattering and Kozai resonances, both coupled with planet-star tidal interactions, have also been suggested as the origin of some close-in planets \citep{1996Sci...274..954R, 1996Natur.384..619W, 2007ApJ...669.1298F}.  \citet{2013ApJ...769...86P} modeled the \kepler~candidate short-period population to investigate possible mechanisms for bringing planets close-in and then halting the inward migration and found a fit for models incorporating the Kozai resonance. Such an origin is problematic for our candidates, though, if they are planets. \citet{2006ApJ...638L..45F} showed that, for planets that begin in highly eccentric orbits, conservation of orbital angular momentum requires that the original pericenter distance for the eccentric orbit is about half the semi-major axis for the final circular orbit of a tidally-evolved planet. Angular momentum is nearly conserved if the tidal evolution is dominated by dissipation within the planet (which is expected for small, rocky planets). Figure \ref{fig:roche_lobe} shows the ratio of the orbital semi-major axes $a$ to the Roche limit $a_\mathrm{Roche}$ for each planet, given by the expression:
\begin{equation}
a_\mathrm{Roche} = 2.44 R_\mathrm{\star} \left(\frac{\rho_\mathrm{\star}}{\rho_\mathrm{p}} \right)^{1/3}
\end{equation}
where $\rho_\mathrm{p}$ is the planetary density. If our candidates are planets with Earth-like densities (5 g/cm$^3$), half their current orbital distances lie inside the Roche limit for many of them (or even inside the star for the KIC 5080636 candidate). On the other hand, \citet{2013ApJ...773L..15R} provided a different Roche lobe relationship for planets with somewhat compressible interiors that suggests the shortest-period candidate orbiting KIC 5080636 might actually require a density $\ge 7$ g/cm$^3$ in order not to be disrupted in its orbit.

\indent As suggested in the introduction, another possible origin for our candidates is as the remnants of disrupted gas giants. Figure \ref{fig:roche_lobe} also shows the relative position of $a/a_\mathrm{Roche}$ if the candidates had densities equal to that of the hot-Jupiter WASP-12 b (for which observations tentatively suggest ongoing disruption -- \citealp{2010ApJ...714L.222F}) and suggests that they would likely have been disrupted with such a low density. Radius inflation, perhaps via ohmic dissipation \citep{2010ApJ...714L.238B}, would enhance the disruption rate. In this origin scenario, the candidates began as hot Jupiters, having arrived at close-in but more-distant-than-at-present orbits in whatever way(s) hot Jupiters do. Then, tidal interactions with the host stars brought them through the hot-Jupiter Roche limit, and the progenitor gas giants were disrupted. If, as suggested by \citet{2013ApJ...773L..15R}, the shortest-period objects orbiting KIC 5080636 does have a density $\ge 7$ g/cm$^3$, this may be further evidence that it is a remnant core.  \citet{Sotin2013_DPS} suggest that the remnant core of a disrupted gas giant might retain an anomalously large density, long after the very high internal pressure is relieved by removal of the atmosphere.

\indent Disruption of close-in gas giants has been investigated by, among others, \citet{2011ApJ...732...74G} and \citet{2013ApJ...762...37L}. For tidal decay to play the role we suggest, it must occur quickly enough to bring the progenitor hot Jupiters to their Roche limits (and possibly to bring the remnant core to a closer-in orbit after that) but slowly enough that we have sufficient time to observe the remnant cores before they are disrupted themselves. Loss of the massive gaseous envelope could slow the tidal decay since the less massive remnant raises smaller tides on the star, but the tidal evolution rate increases so quickly with decreasing orbital separation that it's not clear mass loss would sufficiently stall the orbital decay \citep{2010MNRAS.407..910J}. Indeed, \citet{2012MNRAS.425.2778M} suggest that disruption and accretion of planets may be inevitable, depending on the planet-star density ratio. Future work will investigate this possibility.

\indent Discovery of new extrasolar planets proceeds apace and continues to overturn our understanding of planet formation and evolution. If confirmed as planets, our very short-period candidates add yet another fascinating and unanticipated species to the growing menagerie of planetary systems. Even if they remain unconfirmed, their discovery illustrates how fruitful multiple, complementary analyses of a dataset as rich as \kepler's can be.

\acknowledgments
The authors acknowledge helpful input from Phil Arras, Alan Boss, Joleen Carlberg, Michael Endl, Maki Jackson, Peter McCollough, and Roberto Sanchis-Ojeda. Comments from an anonymous referee also helped enormously.

All of the data presented in this paper were obtained from the Mikulski Archive for Space Telescopes (MAST) and were collected by the \kepler~mission. STScI is operated by the Association of Universities for Research in Astronomy, Inc., under NASA contract NAS5-26555. Support for MAST for non-HST data is provided by the NASA Office of Space Science via grant NNX09AF08G and by other grants and contracts. Funding for the \kepler~mission is provided by the NASA Science Mission directorate. This research also made use of the NASA Exoplanet Archive, which is operated by the California Institute of Technology, under contract with NASA's Exoplanet Exploration Program.

\bibliography{fossil_cores_refs}

\LongTables

\input{make_odd_even_transits_table}

\input{all_binned_photocenter_table}

\input{make_fit_evs_table}

\input{make_derived_params_table}

\end{document}

%% file: make_odd_even_transits_table.tex
\clearpage
\begin{landscape}
\begin{deluxetable*}{ccccccccc}
\tablecaption{Odd/Even Transit Comparison}
\tablehead{\colhead{KIC} & \colhead{trial period} & \colhead{$a$} & \colhead{$b$} & \colhead{$t_\mathrm{1}$} & \colhead{$p_\mathrm{1}$} & \colhead{$t_\mathrm{2}$} & \colhead{$p_\mathrm{2}$} & \colhead{$|\Delta p| / \sigma$} \\
\colhead{} & \colhead{(days)} & \colhead{} & \colhead{} & \colhead{(BJD - 2454833)} & \colhead{} & \colhead{(BJD - 2454833)} & \colhead{} & \colhead{} } 
\startdata
2857722 & 0.420072 &  4.1 $^{+ 0.5}_{- 0.7}$ &  0.3 $^{+ 0.3}_{- 0.2}$ &  352.439 $^{+ 0.001}_{- 0.001}$ &  0.0118 $^{+ 0.0007}_{- 0.0006}$ &  352.6495 $\pm$ 0.0009 &  0.0121 $^{+ 0.0007}_{- 0.0006}$ &   0.35\\
2859299 & 0.321844 &  1.088 $\pm$ 0.007 &  0.996 $^{+ 0.003}_{- 0.002}$ &  131.6747 $\pm$ 0.0005 &  0.023 $^{+ 0.002}_{- 0.001}$ &  131.5135 $\pm$ 0.0008 &  0.018 $^{+ 0.002}_{- 0.001}$ &   2.30\\
3233043 & 0.758912 &  2.27 $\pm$ 0.07 &  0.988 $^{+ 0.008}_{- 0.007}$ &  131.5517 $\pm$ 0.0005 &  0.073 $^{+ 0.005}_{- 0.004}$ &  131.9312 $\pm$ 0.0005 &  0.071 $^{+ 0.005}_{- 0.004}$ &   0.26\\
4175105 & 0.405474 &  2.5 $^{+ 0.4}_{- 0.6}$ &  0.4 $\pm$ 0.3 &  131.804 $\pm$ 0.001 &  0.0102 $^{+ 0.0008}_{- 0.0005}$ &  131.602 $\pm$ 0.001 &  0.0106 $^{+ 0.0008}_{- 0.0005}$ &   0.43\\
4861364 & 0.311636 &  2.6 $^{+ 0.3}_{- 0.5}$ &  0.4 $\pm$ 0.3 &  120.610 $^{+ 0.001}_{- 0.002}$ &  0.0056 $^{+ 0.0004}_{- 0.0003}$ &  120.766 $\pm$ 0.001 &  0.0063 $\pm$ 0.0004 &   1.38\\
5017876 & 0.136038 &  2.1 $^{+ 0.2}_{- 0.3}$ &  0.4 $\pm$ 0.3 &  131.5674 $\pm$ 0.0002 &  0.0152 $^{+ 0.0006}_{- 0.0005}$ &  131.6361 $\pm$ 0.0004 &  0.0109 $^{+ 0.0005}_{- 0.0004}$ &   6.09\\
5080636 & 0.353780 &  3.2 $^{+ 0.7}_{- 0.9}$ &  0.5 $\pm$ 0.3 &  131.7329 $\pm$ 0.0006 &  0.013 $^{+ 0.002}_{- 0.001}$ &  131.5564 $\pm$ 0.0007 &  0.012 $^{+ 0.002}_{- 0.001}$ &   0.20\\
5440651 & 0.502812 &  2.01 $\pm$ 0.03 &  0.998 $\pm$ 0.001 &  131.8613 $\pm$ 0.0003 &  0.102 $\pm$ 0.001 &  131.6099 $\pm$ 0.0002 &  0.116 $\pm$ 0.001 &   8.40\\
5475494 & 0.736180 &  3.7 $^{+ 0.5}_{- 0.6}$ &  0.4 $^{+ 0.2}_{- 0.3}$ &  132.093 $\pm$ 0.001 &  0.0145 $^{+ 0.0008}_{- 0.0006}$ &  131.728 $^{+ 0.004}_{- 0.003}$ &  0.0096 $\pm$ 0.0006 &   5.51\\
5636648 & 0.933442 &  1.53 $^{+ 0.04}_{- 0.03}$ &  0.992 $^{+ 0.006}_{- 0.008}$ &  131.6584 $\pm$ 0.0007 &  0.040 $^{+ 0.003}_{- 0.005}$ &  132.1267 $\pm$ 0.0006 &  0.042 $^{+ 0.003}_{- 0.005}$ &   0.33\\
5896439 & 0.520954 &  1.249 $\pm$ 0.009 &  0.998 $\pm$ 0.002 &  131.9427 $\pm$ 0.0003 &  0.033 $\pm$ 0.001 &  131.6831 $\pm$ 0.0003 &  0.034 $\pm$ 0.001 &   0.93\\
6665064 & 0.698378 &  5.2 $^{+ 0.5}_{- 0.9}$ &  0.4 $^{+ 0.3}_{- 0.2}$ &  132.045 $\pm$ 0.001 &  0.0257 $^{+ 0.0010}_{- 0.0009}$ &  131.6980 $\pm$ 0.0005 &  0.035 $^{+ 0.001}_{- 0.001}$ &   7.04\\
7051984 & 0.678136 &  4.0 $^{+ 0.6}_{- 1.0}$ &  0.5 $\pm$ 0.3 &  121.0213 $^{+ 0.0008}_{- 0.0009}$ &  0.0068 $^{+ 0.0005}_{- 0.0003}$ &  120.6817 $\pm$ 0.0009 &  0.0068 $^{+ 0.0005}_{- 0.0003}$ &   0.03\\
7269881 & 0.613872 &  5 $^{+ 1}_{- 2}$ &  0.5 $^{+ 0.3}_{- 0.4}$ &  131.6381 $\pm$ 0.0009 &  0.011 $^{+ 0.002}_{- 0.001}$ &  131.9446 $^{+ 0.0009}_{- 0.0008}$ &  0.0107 $^{+ 0.0017}_{- 0.0009}$ &   0.13\\
7516809 & 0.483674 &  3.1 $^{+ 0.6}_{- 0.9}$ &  0.4 $^{+ 0.4}_{- 0.3}$ &  131.637 $\pm$ 0.002 &  0.0103 $^{+ 0.0010}_{- 0.0008}$ &  131.878 $\pm$ 0.001 &  0.0132 $^{+ 0.0013}_{- 0.0008}$ &   2.19\\
7582691 & 0.519632 &  7 $^{+ 2}_{- 1}$ &  0.4 $\pm$ 0.3 &  352.442 $\pm$ 0.001 &  0.013 $^{+ 0.002}_{- 0.001}$ &  352.7019 $\pm$ 0.0009 &  0.014 $^{+ 0.002}_{- 0.001}$ &   0.42\\
8260198 & 0.955268 &  1.9 $^{+ 0.4}_{- 0.2}$ &  0.96 $^{+ 0.02}_{- 0.03}$ &  132.1936 $^{+ 0.0008}_{- 0.0007}$ &  0.033 $^{+ 0.004}_{- 0.003}$ &  131.718 $\pm$ 0.003 &  0.020 $\pm$ 0.002 &   3.23\\
8435766 & 0.710012 &  6.2 $^{+ 0.2}_{- 0.4}$ &  0.2 $^{+ 0.2}_{- 0.1}$ &  120.9611 $^{+ 0.0003}_{- 0.0002}$ &  0.0140 $\pm$ 0.0002 &  120.6060 $\pm$ 0.0002 &  0.0140 $\pm$ 0.0002 &   0.19\\
8588377 & 0.994684 &  2.2 $^{+ 0.1}_{- 0.1}$ &  0.963 $^{+ 0.013}_{- 0.008}$ &  131.6208 $\pm$ 0.0004 &  0.050 $^{+ 0.006}_{- 0.003}$ &  132.1186 $\pm$ 0.0007 &  0.037 $^{+ 0.005}_{- 0.002}$ &   2.22\\
8645191 & 0.459470 &  3.2 $\pm$ 0.3 &  0.3 $\pm$ 0.2 &  131.901 $\pm$ 0.001 &  0.0131 $^{+ 0.0005}_{- 0.0004}$ &  131.6693 $^{+ 0.0009}_{- 0.0008}$ &  0.0153 $^{+ 0.0006}_{- 0.0004}$ &   3.28\\
8703491 & 0.616350 &  1.37 $^{+ 0.03}_{- 0.02}$ &  0.995 $\pm$ 0.004 &  120.9869 $\pm$ 0.0004 &  0.026 $^{+ 0.002}_{- 0.003}$ &  120.6785 $\pm$ 0.0004 &  0.025 $^{+ 0.002}_{- 0.003}$ &   0.22\\
9520443 & 0.990182 &  5 $^{+ 1}_{- 1}$ &  0.5 $^{+ 0.4}_{- 0.3}$ &  132.011 $\pm$ 0.001 &  0.018 $^{+ 0.002}_{- 0.001}$ &  131.516 $\pm$ 0.002 &  0.015 $^{+ 0.002}_{- 0.001}$ &   0.93\\
9597729 & 0.876318 &  5.3 $^{+ 0.8}_{- 1.1}$ &  0.4 $\pm$ 0.3 &  132.386 $\pm$ 0.002 &  0.0101 $\pm$ 0.0006 &  131.949 $\pm$ 0.001 &  0.0107 $^{+ 0.0006}_{- 0.0005}$ &   0.70\\
9752973 & 0.477818 &  3.7 $^{+ 0.7}_{- 0.8}$ &  0.6 $^{+ 0.2}_{- 0.3}$ &  120.5700 $\pm$ 0.0009 &  0.0038 $^{+ 0.0003}_{- 0.0002}$ &  120.807 $\pm$ 0.001 &  0.0034 $^{+ 0.0003}_{- 0.0002}$ &   1.19\\
9883561 & 0.349676 &  2.9 $\pm$ 0.4 &  0.5 $^{+ 0.2}_{- 0.3}$ &  131.8443 $\pm$ 0.0006 &  0.0156 $^{+ 0.0009}_{- 0.0007}$ &  131.6685 $\pm$ 0.0005 &  0.0171 $^{+ 0.0009}_{- 0.0008}$ &   1.31\\
9943435 & 0.777686 &  4.1 $^{+ 0.7}_{- 1.1}$ &  0.5 $\pm$ 0.3 &  132.288 $\pm$ 0.001 &  0.0115 $^{+ 0.0009}_{- 0.0006}$ &  131.900 $\pm$ 0.002 &  0.0075 $^{+ 0.0006}_{- 0.0005}$ &   4.63\\
10402660 & 0.488184 &  1.64 $\pm$ 0.04 &  0.997 $\pm$ 0.002 &  352.6255 $\pm$ 0.0004 &  0.055 $\pm$ 0.002 &  352.3815 $\pm$ 0.0004 &  0.052 $^{+ 0.001}_{- 0.002}$ &   1.20\\
10453521 & 0.896552 &  1.55 $\pm$ 0.06 &  0.990 $^{+ 0.006}_{- 0.007}$ &  907.504 $\pm$ 0.001 &  0.034 $\pm$ 0.004 &  907.057 $\pm$ 0.001 &  0.035 $\pm$ 0.004 &   0.28\\
11496490 & 0.414110 &  5.0 $\pm$ 0.5 &  0.3 $\pm$ 0.2 &  131.7289 $\pm$ 0.0007 &  0.0094 $\pm$ 0.0004 &  131.5219 $\pm$ 0.0007 &  0.0091 $^{+ 0.0004}_{- 0.0003}$ &   0.64\\
11709423 & 0.768954 &  5.2 $^{+ 0.8}_{- 1.8}$ &  0.4 $^{+ 0.4}_{- 0.3}$ &  352.4154 $\pm$ 0.0005 &  0.025 $^{+ 0.003}_{- 0.001}$ &  352.8019 $\pm$ 0.0008 &  0.020 $^{+ 0.002}_{- 0.001}$ &   1.85\\
11969092 & 0.425642 &  3.1 $^{+ 0.4}_{- 0.7}$ &  0.4 $\pm$ 0.3 &  131.572 $\pm$ 0.001 &  0.0116 $^{+ 0.0010}_{- 0.0007}$ &  131.784 $\pm$ 0.001 &  0.0118 $^{+ 0.0010}_{- 0.0007}$ &   0.18\\
11972387 & 0.327894 &  3.3 $^{+ 0.5}_{- 0.7}$ &  0.5 $\pm$ 0.3 &  120.7921 $\pm$ 0.0002 &  0.0157 $^{+ 0.0009}_{- 0.0006}$ &  120.6286 $\pm$ 0.0002 &  0.0152 $^{+ 0.0009}_{- 0.0006}$ &   0.49\\
12023078 & 0.623436 &  4.1 $^{+ 0.5}_{- 0.8}$ &  0.5 $\pm$ 0.2 &  121.0838 $\pm$ 0.0009 &  0.0084 $^{+ 0.0005}_{- 0.0003}$ &  120.7721 $\pm$ 0.0006 &  0.0101 $^{+ 0.0006}_{- 0.0004}$ &   2.84\\
12120286 & 0.356664 &  2.3 $\pm$ 0.3 &  0.4 $\pm$ 0.3 &  131.713 $\pm$ 0.001 &  0.0120 $^{+ 0.0005}_{- 0.0004}$ &  131.537 $\pm$ 0.001 &  0.0122 $^{+ 0.0005}_{- 0.0004}$ &   0.40\\
\enddata
\tablecomments{The column labeled ``trial period'' shows the period on which the data were folded, $a$ the orbital semi-major axis scaled to the sum of radii, $b$ the impact parameter, $t_\mathrm{1/2}$ the mid-transit time for the first/second transit, $p_\mathrm{1/2}$ the radius ratio for the first/second transit, and $|\Delta p/\sigma|$ the difference between radius ratios divided by their mutual uncertainties.}
\label{tbl:check_odd_even_transits}
\end{deluxetable*}
\clearpage
\end{landscape}

%% file: all_binned_photocenter_table.tex
\begin{deluxetable}{cccc}
\tablecaption{Scaled Photocenter Shifts}
\tablehead{\colhead{KIC} & \colhead{trial period} & \colhead{$|\Delta \alpha|/\sigma$} & \colhead{$|\Delta \delta|/\sigma$} \\
\colhead{} & \colhead{(days)} & \colhead{} & \colhead{}}
\startdata
2857722 & 0.420072 / 0.210036 & 3.78 / 6.64 & 11.54 / 14.88 \\
2859299 & 0.321844 & 39.80 & 58.33 \\
3233043 & 0.379456 & 14.05 & 5.80 \\
4175105 & 0.202737 & 4.02 & 48.96 \\
4861364 & 0.311636 / 0.155818 & 13.63 / 15.44 & 8.07 / 7.79 \\
5017876 & 0.136038 & 10.53 & 1.96 \\
\textbf{5080636} & \textbf{0.353780 / 0.176890} & \textbf{0.03 / 0.84} & \textbf{0.58 / 1.45} \\
5440651 & 0.251406 & 67.89 & 11.37 \\
5475494 & 0.736180 & 14.88 & 14.61 \\
5636648 & 0.466721 & 91.02 & 12.60 \\
5896439 & 0.260477 & 141.76 & 161.97 \\
6665064 & 0.698378 / 0.349189 & 26.49 / 26.27 & 12.42 / 12.98 \\
7051984 & 0.339068 & 3.02 & 1.07 \\
\textbf{7269881} & \textbf{0.306936} & \textbf{0.33} & \textbf{0.59} \\
7516809 & 0.483674 / 0.241837 & 1.93 / 2.40 & 46.68 / 47.36 \\
7582691 & 0.259816 & 4.23 & 2.14 \\
8260198 & 0.955268 & 44.13 & 18.88 \\
\textbf{8435766} & \textbf{0.355006} & \textbf{0.11} & \textbf{2.56} \\
8588377 & 0.994684 / 0.497342 & 72.88 / 90.76 & 85.46 / 117.94 \\
8645191 & 0.459470 / 0.229735 & 4.72 / 5.72 & 25.95 / 27.57 \\
8703491 & 0.308175 & 180.77 & 201.47 \\
9520443 & 0.990182 & 71.68 & 53.20 \\
9597729 & 0.876318 / 0.438159 & 34.32 / 49.97 & 47.93 / 70.74 \\
9752973 & 0.477818 & 10.16 & 8.51 \\
9883561 & 0.349676 / 0.174838 & 107.91 / 135.72 & 72.28 / 85.94 \\
9943435 & 0.777686 & 114.57 & 85.20 \\
10402660 & 0.244092 & 26.68 & 33.02 \\
\textbf{10453521} & \textbf{0.448276} & \textbf{1.16} & \textbf{2.41} \\
11496490 & 0.414110 / 0.207055 & 4.69 / 6.35 & 2.49 / 2.30 \\
11709423 & 0.768954 / 0.384477 & 63.49 / 86.60 & 68.59 / 78.44 \\
11969092 & 0.212821 & 2.65 & 3.17 \\
11972387 & 0.163947 & 8.72 & 1.63 \\
12023078 & 0.623436 / 0.311718 & 165.55 / 224.29 & 80.92 / 90.75 \\
12120286 & 0.178332 & 38.39 & 5.25 \\
\enddata
\tablecomments{The column labeled KIC show the KIC numbers, ``trial period'' the orbital period considered, and $\Delta \alpha/\sigma$ and $\Delta \delta/\sigma$ the absolute in-transit displacement in RA and Dec respectively (for the corresponding trial period) and normalized to the uncertainty. Those candidates with normalized displacements $ < 3$ are highlighted with bold text.}
\label{tbl:all_binned_photocenter_table}
\end{deluxetable}

%% file: make_fit_evs_table.tex
\clearpage
\begin{landscape}
\begin{deluxetable}{cccccccccc}
\tablecaption{Photometric Model Parameters}
\tablehead{\colhead{KIC} & \colhead{period} & \colhead{$a$} & \colhead{$t_\mathrm{0}$} & \colhead{$p$} & \colhead{$b$} & \colhead{$A_\mathrm{DB}$} & \colhead{$A_\mathrm{ELV}$} & \colhead{$A_\mathrm{ill}$} & \colhead{$D$} \\
\colhead{} & \colhead{(days)} & \colhead{} & \colhead{(BJD - 2454833)} & \colhead{} & \colhead{} & \colhead{(ppm)} & \colhead{(ppm)} & \colhead{(ppm)} & \colhead{(ppm)}}
\startdata
5080636 &  0.176896 $\pm$ 0.000002 &  1.9 $^{+ 0.8}_{- 0.7}$ &  131.5395 $^{+ 0.0009}_{- 0.0011}$ &  0.0085 $\pm$ 0.0009 &  0.5 $\pm$ 0.3 &  4 $^{+ 3}_{- 2}$ &  10 $^{+ 5}_{- 6}$ &  26 $\pm$ 6 &  8 $^{+ 8}_{- 6}$ \\ 
7269881 &  0.306936 $\pm$ 0.000006 &  4.0 $\pm$ 1.0 &  131.6374 $\pm$ 0.0006 &  0.0109 $^{+ 0.0013}_{- 0.0010}$ &  0.4 $\pm$ 0.3 &  0.7 $^{+ 0.8}_{- 0.6}$ &  2 $\pm$ 2 &  5 $\pm$ 2 &  14 $\pm$ 7 \\ 
8435766 &  0.3550080 $\pm$ 0.0000006 &  3.2 $^{+ 0.2}_{- 0.3}$ &  120.60341 $^{+ 0.00010}_{- 0.00009}$ &  0.0140 $^{+ 0.0004}_{- 0.0002}$ &  0.3 $\pm$ 0.2 &  0.5 $^{+ 0.4}_{- 0.3}$ &  0.4 $^{+ 0.4}_{- 0.3}$ &  4.2 $\pm$ 0.7 &  9 $\pm$ 1 \\ 
10453521 &  0.44841 $\pm$ 0.00003 &  2.1 $^{+ 0.8}_{- 0.7}$ &  907.0429 $^{+ 0.0004}_{- 0.0005}$ &  0.027 $^{+ 0.005}_{- 0.003}$ &  0.9 $^{+ 0.1}_{- 0.2}$ &  18 $\pm$ 5 &  9 $^{+ 6}_{- 5}$ &  30 $\pm$ 6 &  16.0 $^{+ 13.6}_{- 11.3}$ \\ 
\enddata
\tablecomments{The column labeled ``period'' shows the best-fit period and uncertainties, $a$ the orbital semi-major axis scaled to the stellar radius, $t_\mathrm{0}$ the mid-transit time for the first transit observed, $p$ the radius ratio, $b$ the impact parameter, $A_\mathrm{DB}$ the amplitude of the Doppler beaming signal, $A_\mathrm{ELV}$ the amplitude of the ellipsoidal variation, $A_\mathrm{ill}$ the amplitude of the planetary phase curve, and $D$ the eclipse depth.}
\label{tbl:fit_evs}
\end{deluxetable}
\clearpage
\end{landscape}

%% file: make_derived_params_table.tex
\begin{deluxetable*}{cccccccccc}
\tablecaption{Derived Physical Parameters}
\tablehead{\colhead{KIC} & \colhead{$u$} & \colhead{$g$} & \colhead{$R_\mathrm{p}$} & \colhead{$T_\mathrm{p, eq}$} & \colhead{$T_\mathrm{p, meas}$} & \colhead{$q_\mathrm{DB}$} & \colhead{$M_\mathrm{p, DB}$} & \colhead{$q_\mathrm{ELV}$} & \colhead{$M_\mathrm{p, ELV}$} \\
\colhead{} & \colhead{} & \colhead{} & \colhead{($R_\mathrm{Earth}$)} & \colhead{(K)} & \colhead{(K)} & \colhead{(ppm)} & \colhead{($M_\mathrm{Earth}$)} & \colhead{(ppm)} & \colhead{($M_\mathrm{Earth}$)}}
\startdata
5080636 & 0.705022 & 0.476015 &  0.58 $\pm$ 0.06 &  2200 $\pm$ 600 &  2600 $\pm$ 400 &  500 $\pm$ 500 &  80 $\pm$ 70 &  50 $\pm$ 90 &  8 $\pm$ 14 \\ 
7269881 & 0.681427 & 0.510710 &  0.95 $^{+ 0.11}_{- 0.09}$ &  2100 $\pm$ 400 &  3300 $\pm$ 400 &  100 $\pm$ 200 &  30 $\pm$ 50 &  80 $\pm$ 150 &  20 $\pm$ 40 \\ 
8435766 & 0.684163 & 0.510355 &  1.44 $^{+ 0.04}_{- 0.02}$ &  2300 $\pm$ 100 &  2870 $\pm$ 70 &  90 $\pm$ 90 &  30 $\pm$ 30 &  9 $\pm$ 11 &  3 $\pm$ 3 \\ 
10453521 & 0.514518 & 0.257134 &  3.8 $^{+ 0.7}_{- 0.4}$ &  3800 $\pm$ 1000 &  3100 $\pm$ 400 &  5000 $\pm$ 2000 &  1900 $\pm$ 800 &  90 $\pm$ 150 &  30 $\pm$ 60 \\ 
\enddata
\tablecomments{The column $u$ shows the linear limb-darkening coefficient, $g$ the gravity-darkening exponent, $R_\mathrm{p}$ the planetary radius, $T_\mathrm{p, eq}$ the day side blackbody temperature expected for radiative equilibrium, $T_\mathrm{p, meas}$ the effective blackbody temperature estimated from the eclipse depth, $q_\mathrm{DB/ELV}$ the mass ratio estimate from the Doppler beaming/ellipsoidal variation signal, and $M_\mathrm{p, DB/ELV}$ the corresponding planetary mass. For most cases, the Doppler beaming/ellipsoidal variation signal are not robustly detected, and so the corresponding upper limits on the mass values should be interpreted as upper limits for non-detections. Uncertainties in the $T_\mathrm{p, eq}$ column and rightward have been approximated as Gaussian.}
\label{tbl:derived_params}
\end{deluxetable*}